\documentclass[preprint,showpacs,preprintnumbers,amsmath,amssymb,superscriptaddress,11pt,graphicx]{revtex4}

\usepackage{graphicx}
\usepackage{dcolumn}
\usepackage{bm}
\usepackage{amssymb}

\newcommand{\bea}{\begin{eqnarray}}
\newcommand{\ena}{\end{eqnarray}}

\begin{document}

\title{Separating E and B types of polarization on an incomplete sky}

\author{Wen Zhao}
\email{Wen.Zhao@astro.cf.ac.uk} \affiliation{School of Physics and
Astronomy, Cardiff University, Cardiff, CF24 3AA, U.K.}\affiliation{Wales Institute of Mathematical and
Computational Sciences, Swansea, SA2 8PP, U.K.}
\affiliation{Department of Physics, Zhejiang University of
Technology, Hangzhou, 310014, P.R.China}

\author{Deepak Baskaran}
\email{Deepak.Baskaran@astro.cf.ac.uk} \affiliation{School of
Physics and Astronomy, Cardiff University, Cardiff, CF24 3AA,
U.K.}\affiliation{Wales Institute of Mathematical and
Computational Sciences, Swansea, SA2 8PP, U.K.}

\date{\today}


\begin{abstract}

Detection of magnetic-type ($B$-type) polarization in the Cosmic Microwave Background (CMB) radiation plays a crucial role in probing the relic gravitational wave (RGW) background. In this paper, we propose a new method to deconstruct a polarization map on an incomplete sky in real space into purely electric and magnetic polarization type maps, ${\mathcal{E}}(\hat{\gamma})$ and ${\mathcal{B}}(\hat{\gamma})$, respectively. The main properties of our approach are as follows: 
Firstly, the fields  ${\mathcal{E}}(\hat{\gamma})$ and ${\mathcal{B}}(\hat{\gamma})$ are constructed in real space with a minimal loss of information. This loss of information arises due to the removal of a narrow edge of the constructed map in order to remove various numerical errors, including those arising from finite pixel size.
Secondly, this method is fast and can be efficiently applied to high resolution maps due to the use of the fast spherical harmonics transformation. Thirdly, the constructed fields, ${\mathcal{E}}(\hat{\gamma})$ and ${\mathcal{B}}(\hat{\gamma})$, are scalar fields. For this reason various techniques developed to deal with temperature anisotropy maps can be directly applied to analyze these fields. As a concrete example, we construct and analyze an unbiased estimator for the power spectrum of the $B$-mode of polarization $C_{\ell}^{BB}$. Basing our results on the performance of this estimator, we discuss the RGW detection ability of two future ground-based CMB experiments, QUIET and POLARBEAR.

\end{abstract}



\maketitle



\section{Introduction \label{section1}}

The extreme conditions in the very early Universe produce primordial perturbations
of two generic types, namely, density perturbations (scalar perturbations) and
relic gravitational waves (tensor perturbations) \cite{gr74,guth}. In the simplest scenarios,
these perturbations are characterized by a nearly scale invariant primordial power 
spectra. The experimental determination of the parameters specifying these power spectra provides
an important method to investigating the physics of the very early Universe.
The Cosmic Microwave Background Radiation (CMB) has proved to be a valuable
tool in this respect. Scalar and tensor perturbations leave an observable imprint
in the temperature and polarization anisotropies of the CMB. The recent experimental
effort, including the WMAP satellite \cite{wmap}, QUaD \cite{quad}, BICEP \cite{bicep} and so on \cite{dasi,cbi,capmap}\cite{boomerang,maxipol},
has lead to a robust determination of the parameters characterizing the primordial density perturbations.
On the other hand, the detection of relic gravitational waves (RGWs) remains an outstanding
experimental challenge, and a key task for current, upcoming and future planned CMB observations
on the ground \cite{keating-ground,quad,bicep,clover,POLARBEAR,quiet,quijote,sptpol,actpol}, 
on balloons \cite{ebex,pappa,spider} and in the space \cite{taskforce,planck,b-pol,litebird,cmbpol}.

Both density perturbations and RGWs contribute to the various
CMB anisotropy power spectra, namely $TT$ (temperature), $EE$ (``electric"-type polarization) and
$TE$ (temperature-polarization cross correlation) 
\cite{cmb1,hucmb,ttee1,ttee2,ttee3,ttee4,ttee5,ttee6,polnarev,ttteee}. In addition,
RGWs produce the ``magnetic"-type ($BB$) polarization that is not
produced by density perturbations \cite{bb}.
In principle, all of these information channels ($TT$, $TE$, $EE$ and $BB$) shoul be used to infer
the RGW signal in the CMB. However, if the contribution of RGWs to the CMB
is small ($r\lesssim0.05$), the $BB$ channel will the best venue for detecting RGWs \cite{ttteee}.  

The separation of the polarization field into electric and magnetic components is
a subtle issue. In practice, in a CMB experiment, one directly observes only the 
Stokes polarization components $Q$ and $U$. Given a full sky map of the components
$Q$ and $U$, one can construct the so-called $E$-mode and $B$-mode of polarization 
(sometimes referred to as $G$ and $C$ modes, respectively) using spin-spherical
harmonics expansion in an unambiguous manner. It is important to keep in mind that,
by virtue of construction, $E$ and $B$ modes of polarization are non-local quantities,
and require the information on Stokes parameters on the complete sky. However, in realistic cases 
(ground-based, balloon-borne and satellite experiments), the Stokes parameters
are measured only on a fractional portion of the sky. In this situation, the simplest
method for constructing electric and magnetic polarization fields, using the 
spin-spherical harmonics leads to mutual contamination, often referred to as the $EB$-mixing.
This $EB$-mixing can become a dominant hinderance for detecting the RGW signal \cite{challinor}.
In order to overcome this difficulty, numerous methods have been developed to
separate $E$ and $B$ types of polarization on an incomplete sky \cite{lewis,pixel,smith,smith2,needlet,cao,kim}.
However, these methods suffer from one or several of the following drawbacks -
they are slow in practice, they are difficult to realize in pixel space, and/or they lead to partial information loss.

In the current work, we develop a novel method to separate the electric and magnetic components
of polarization on a partial sky. In contrast to previous works \cite{lewis,pixel,smith}, in this paper
we focus mainly on the construction of pure $E$ and $B$ types of polarization in the real space,
as opposed to constructions in harmonics space. Our method is based on a simple redefinition of electric
and magnetic components of polarization, so as to make them only depending on the differential of the $Q$ and $U$ fields.
On a full sky, this definition is equivalent to the standard definition. However, due to the differential nature, our definition is
directly extendable to polarization maps given on an incomplete sky. The main advantages of our
method are as follows. Firstly, the loss of information is small, and is only caused by the removal of
a narrow edge around the observed portion of the sky to reduce numerical errors. Secondly,
this method is easy to realize for pixelized polarization maps, and is sufficiently fast so as to be practical for
high resolution full sky surveys. Thirdly, the method leads to construction of scalar $E$ and $B$ type fields. For
this reason, one can directly apply the various techniques developed for temperature anisotropy. In particular,
along this route, we construct an unbiased estimator for the $B$-mode of polarization power spectrum and
gauge its performance in detecting RGWs.

The outline of the paper is as follows. In Sec.~\ref{section2}, we introduce the basic notations and explain
the general ideas behind our method. In this section we derive the basic equations, and construct the pure
electric and magnetic fields ${\mathcal{E}}(\hat{\gamma})$ and ${\mathcal{B}}(\hat{\gamma})$, given the polarization
map ($Q(\hat{\gamma})$, $U(\hat{\gamma})$) on an incomplete sky. In Sec.~\ref{section3}, we apply this method
in practice for a pixelized polarization map. In this section we explicitly construct the ${\mathcal{B}}(\hat{\gamma})$ field, and discuss the various
sources of contamination. We go on to discuss the effect edge removal, size of pixels and experimental beam size on
the resulting contaminations. In Sec.~\ref{section4}, we focus on the applying our method to small sky surveys. Combining
our method with the pseudo-$C_{\ell}$ estimator method we construct an unbiased estimator for the $B$-mode power spectrum.
We show that, in practice, our estimator performs only slightly worse than an estimator in an idealized situation with no loss of information.
Based on this estimator, we analyze the ability to detect relic gravitational waves through their signature in $B$-mode of polarization in
two planned ground-based CMB experiments, QUIET and POLARBEAR. We conclude in Sec.~\ref{section5} with a brief summary of our
main results.


\section{Decomposition of the polarization field into electric and magnetic components on an incomplete sky \label{section2}}

Let us first give a brief recount of the standard procedure to construct
electric $E$-mode and magnetic $B$-mode of polarization, given a complete sky.
For mathematical simplicity, it is convenient to introduce the complex conjugate
polarization fields $P_{\pm}$ as follows
\bea
P_{\pm}(\hat{\gamma}) \equiv Q(\hat{\gamma}) \pm i U(\hat{\gamma}),
\label{p+-1}
\ena
where $\hat{\gamma}$ denotes the position on the sky, 
and $Q$ and $U$ are assumed to be real fields on the sky. The fields $P_{\pm}$,
being $\pm2$ spin-weighted quantities, can be expanded over appropriate
spin-harmonic functions (see \cite{cmb2} for instance):
\bea
P_{\pm}(\hat{\gamma})
=\sum_{\ell m} a_{\pm 2,\ell m}  ~_{\pm 2}Y_{\ell m}(\hat{\gamma}), 
\label{p+}
\label{p-} 
\ena
where $~_{\pm2}Y_{\ell m}(\hat{\gamma})$ are the spin-weighted
spherical harmonics. The multipole coefficients $a_{\pm2,\ell m}$
are calculable by
\bea
a_{\pm2,\ell m} = \int d\hat{\gamma} P_{\pm}(\hat{\gamma})~_{\pm 2}Y_{\ell m}^*(\hat{\gamma}).
\label{multipoledecomposition}
\ena
The $E$ and $B$ mode multipoles are defined
in terms of the coefficients $a_{\pm2,\ell m}$ in the following manner
\bea 
E_{\ell m}\equiv-\frac{1}{2}\left[a_{2,\ell m}+a_{-2,\ell m}\right],
~~B_{\ell m}\equiv-\frac{1}{2i}\left[a_{2,\ell m}-a_{-2,\ell
m}\right]. \label{eblm}
\ena 
One can now define the electric polarization sky map $E(\hat{\gamma})$,
and the magnetic polarization sky map $B(\hat{\gamma})$ as
\bea
E(\hat{\gamma})\equiv\sum_{\ell m}   E_{\ell m} Y_{\ell
m}(\hat{\gamma}), ~~ B(\hat{\gamma})\equiv\sum_{\ell m} B_{\ell m}
Y_{\ell m}(\hat{\gamma}). 
\label{eb-definition}
\ena
The power spectra of $E$ and $B$ modes of polarization are defined, in terms of
the multipole coefficients $E_{\ell m}$ and $B_{\ell m}$, as
\bea
C_{\ell}^{EE}\equiv \frac{1}{2\ell+1}\sum_{m} \langle E_{\ell m} E^*_{\ell m}\rangle,~~
C_{\ell}^{BB}\equiv  \frac{1}{2\ell+1}\sum_{m} \langle B_{\ell m} B^*_{\ell m}\rangle,
\label{ebpowerspectrum}
\ena
where the brackets denote the average over all realizations.

It is important to note that, the polarization sky maps $E(\hat{\gamma})$ and 
$B(\hat{\gamma})$, are constructed out of underlying $Q(\hat{\gamma})$ and
$U(\hat{\gamma})$ maps in a non local manner. This is to say that, the value of 
the $E$ or $B$ field at a given point $\hat{\gamma}$, in virtue 
of (\ref{eb-definition}), depend on multipole coefficients $E_{\ell m}$ and $B_{\ell m}$,
respectively. These coefficients, in turn, depend on integral of $P_{\pm}(\hat{\gamma})$ over
the full sky (see (\ref{multipoledecomposition}) and (\ref{eblm})). Therefore, one requires the 
knowledge of $Q(\hat{\gamma})$ and $U(\hat{\gamma})$ (or equivalently $P_{\pm}(\hat{\gamma})$)
over the entire sky in order to construct the $E(\hat{\gamma})$ and $B(\hat{\gamma})$ fields.

As was mentioned previously, in realistic scenarios, one does not have information
on $Q$ and $U$ fields on the entire sky. For this reason (\ref{multipoledecomposition}), (\ref{eblm})
and (\ref{eb-definition}) cannot be applied directly to construct $E$ and $B$ types
of polarization maps on an incomplete sky. In order to avoid this problem, in the 
present paper we adopt a different but related definition for electric and magnetic
polarization maps
\bea
{\mathcal{E}}(\hat{\gamma}) \equiv 
-\frac{1}{2}\left[\bar{\eth}_1\bar{\eth}_2 P_{+}(\hat{\gamma})+\eth_1\eth_2 P_{-}(\hat{\gamma})\right] ,
\label{epsilon}\\
{\mathcal{B}}(\hat{\gamma}) \equiv
-\frac{1}{2i}\left[\bar{\eth}_1\bar{\eth}_2 P_{+}(\hat{\gamma})-\eth_1\eth_2 P_{-}(\hat{\gamma})\right] , 
\label{beta}
\ena 
where $\bar{\eth}_s$ and ${\eth}_s$ ($s=1,2$) are the spin lowering and 
raising operators, respectively
 \bea
 \bar{\eth}_s\equiv -(\sin\theta)^{-s}
 \left\{\frac{\partial }{\partial \theta}
 -\frac{i}{\sin\theta}\frac{\partial }{\partial
 \phi}\right\}\left[\sin^s\theta\right], \label{eth1} \\
 {\eth}_s\equiv -(\sin\theta)^{-s}
 \left\{\frac{\partial }{\partial \theta}
 +\frac{i}{\sin\theta}\frac{\partial }{\partial
 \phi}\right\}\left[\sin^s\theta\right]. \label{eth2}
 \ena
The definitions (\ref{epsilon}) and (\ref{beta}) for ${\mathcal{E}}$ and ${\mathcal{B}}$ 
have been previously discussed in the literature
\cite{zaldarriaga,cmb3}\cite{ttee3}\cite{smith2,needlet,kim}\cite{baumann}. 
These have often been denoted as $\tilde E$ and $\tilde B$, respectively (see for example \cite{zaldarriaga,baumann}). The
fields ${\mathcal{E}}$ and ${\mathcal{B}}$ are equivalent to the fields $E$ and $B$ 
introduced in \cite{ttee3}, where these fields were introduced as two independent invariants
constructed out of the second covariant derivatives polarization tensor (see Eq.~(36) in \cite{ttee3}).
In the present paper, to maintain a clear distinction from ($E$, $B$) in (\ref{eb-definition}), we shall 
use the (${\mathcal{E}}$, ${\mathcal{B}}$) notation. 

The constructed electric and magnetic fields are scalar fields on the sphere. Therefore,
assuming ${\mathcal{E}}$ and ${\mathcal{B}}$ given on a full sky, one can determine the
spherical harmonics decomposition coefficients
\bea 
{\mathcal{E}}_{\ell m}\equiv \int  d\hat{\gamma}~
{\mathcal{E}}(\hat{\gamma}) Y_{\ell m}^*(\hat{\gamma}),~~
{\mathcal{B}}_{\ell m}\equiv \int 
 d\hat{\gamma}~ {\mathcal{B}}(\hat{\gamma}) Y_{\ell
m}^*(\hat{\gamma}). 
\label{mathcalelmblm}
\ena
These relations can be inverted to give
\bea
{\mathcal{E}}(\hat{\gamma}) = \sum_{\ell m} {\mathcal{E}}_{\ell m}Y_{\ell m}(\hat{\gamma}),
~~
{\mathcal{B}}(\hat{\gamma}) = \sum_{\ell m} {\mathcal{B}}_{\ell m}Y_{\ell m}(\hat{\gamma}).
\label{mathcalelmblm2}
\ena
The multipole coefficients ${\mathcal{E}}_{\ell m}$ and ${\mathcal{B}}_{\ell m}$ 
are related to $E_{\ell m}$ and ${{B}}_{\ell m}$ defined in (\ref{eblm}) by a 
$\ell$-dependent numerical factor $N_{\ell}\equiv\sqrt{(\ell+2)!/(\ell-2)!}$ \cite{zaldarriaga,cmb3,ttee3}
\bea
 {\mathcal{E}}_{\ell m} = N_{\ell} E_{\ell m}, ~~{\mathcal{B}}_{\ell m} = N_{\ell} B_{\ell
 m}.
 \label{eb-mathcaleb}
\ena
One can also define the power spectra of ${\mathcal{E}}$ and ${\mathcal{B}}$ modes of polarization 
\bea
C_{\ell}^{{\mathcal{E}}{\mathcal{E}}}\equiv  \frac{1}{2\ell+1}\sum_{m} \langle{\mathcal{E}}_{\ell m} {\mathcal{E}}^*_{\ell m}\rangle,~~
C_{\ell}^{{\mathcal{B}}{\mathcal{B}}}\equiv  \frac{1}{2\ell+1}\sum_{m} \langle{\mathcal{B}}_{\ell m} {\mathcal{B}}^*_{\ell m}\rangle.
\ena
These are related with the power spectra $C_{\ell}^{EE}$ and $C_{\ell}^{BB}$ through relations 
\bea
C_{\ell}^{{\mathcal{E}}{\mathcal{E}}} = N_{\ell}^2 C_{\ell}^{EE},~~C_{\ell}^{{\mathcal{B}}{\mathcal{B}}} = N_{\ell}^2 C_{\ell}^{BB}. \label{new-old-relations}
\ena
Thus, in comparison with $C_{\ell}^{EE}$ and $C_{\ell}^{BB}$, the power spectra 
$C_{\ell}^{{\mathcal{E}}{\mathcal{E}}}$ and $C_{\ell}^{{\mathcal{B}}{\mathcal{B}}}$ are ``bluer", due to the factor $N_{\ell}^2$. Note that, the 
relations (\ref{eb-mathcaleb}) and (\ref{new-old-relations}) assume the polarization fields to be given on a complete sky.

It is important to point out that, the quantities ${\mathcal{E}}$ and
${\mathcal{B}}$ defined in (\ref{epsilon}) and (\ref{beta}) only depend on the differential of the $Q$ and $U$ fields
by construction. Therefore, these definitions can be, in principle,
applied in the case of $Q$ and $U$ given on an incomplete portion of the sky, to construct
the ${\mathcal{E}}$ and ${\mathcal{B}}$ fields on this portion. We now proceed to 
discuss the relevant steps for this construction on an incomplete sky.

In order to describe the partial sky observations, we firstly
introduce the mask window function $W(\hat{\gamma})$. This mask function is
non-zero only in the observational region of the sky. 
In addition, we shall assume that $W(\hat{\gamma})$ is a real function.
In particular, the special case with
$W(\hat{\gamma})=1$ in the observational region corresponds to the widely
discussed top-hat window function. In the present paper, we denote this special case
of a top-hat window function as $w(\hat{\gamma})$. With the introduction of the window function
$W(\hat{\gamma})$, the analysis of the polarization field $P_{\pm}(\hat{\gamma})$ defined
on the partial region of the sky becomes equivalent to studying the masked field 
$P_{\pm}(\hat{\gamma})W(\hat{\gamma})$ defined on the complete sky.

In the general case of an arbitrary mask, one can define two full sky maps 
$\tilde{\mathcal{E}}(\hat{\gamma})$ and $\tilde {\mathcal{B}}(\hat{\gamma})$
constructed out of observational data
\bea 
\tilde{{\mathcal{E}}}(\hat{\gamma})\equiv
 -\frac{1}{2}\left[\bar{\eth}_1\bar{\eth}_2
 (P_{+}(\hat{\gamma})W(\hat{\gamma}))
 +\eth_1\eth_2 (P_{-}(\hat{\gamma})W(\hat{\gamma}))\right],
 \label{pseudo-epsilon}\\
 \tilde{{\mathcal{B}}}(\hat{\gamma}) \equiv
 -\frac{1}{2i}\left[\bar{\eth}_1\bar{\eth}_2 (P_{+}(\hat{\gamma})W(\hat{\gamma}))-\eth_1\eth_2
 (P_{-}(\hat{\gamma})W(\hat{\gamma}))\right]. 
\label{pseudo-beta}
\ena
Due to the presence of the window function $W(\hat{\gamma})$,
the two maps $\tilde{\mathcal{E}}(\hat{\gamma})$ and $\tilde {\mathcal{B}}(\hat{\gamma})$
do not correspond to pure electric and magnetic types of polarization.
The main task of this work is to construct pure ${\mathcal{E}}(\hat{\gamma})$ 
and ${\mathcal{B}}(\hat{\gamma})$ fields out of $\tilde{\mathcal{E}}(\hat{\gamma})$ 
and $\tilde {\mathcal{B}}(\hat{\gamma})$. 

Moving on, we define the multipole decomposition coefficients $\tilde{{E}}_{\ell m}$ and $\tilde{{B}}_{\ell m}$ as
\bea
\tilde{{E}}_{\ell m} = \frac{1}{N_\ell} \int d\hat{\gamma}~ \tilde{{\mathcal{E}}}(\hat{\gamma})~Y_{\ell m}^*(\hat{\gamma}),~~
\tilde{{B}}_{\ell m} = \frac{1}{N_\ell} \int d\hat{\gamma}~ \tilde{{\mathcal{B}}}(\hat{\gamma})~Y_{\ell m}^*(\hat{\gamma}).
\ena
With this definition, the $\tilde{\mathcal{E}}$ and 
$\tilde {\mathcal{B}}$ fields can be expanded in terms of the
multipole coefficients $\tilde{{E}}_{\ell m}$ and $\tilde{{B}}_{\ell m}$ in the following manner
\bea
\tilde{{\mathcal{E}}}(\hat{\gamma}) = \sum_{\ell m} N_\ell \tilde{{E}}_{\ell m} Y_{\ell m}(\hat{\gamma}),~~
\tilde{{\mathcal{B}}}(\hat{\gamma}) = \sum_{\ell m} N_\ell \tilde{{B}}_{\ell m} Y_{\ell m}(\hat{\gamma}).
\label{pseudo-epsilonbeta}
\ena
The multipole decomposition coefficients $\tilde{{E}}_{\ell m}$ and $\tilde{{B}}_{\ell m}$ can be calculated
in an alternative manner. One can begin by defining the complex polarization fields 
$\tilde{P}_{\pm}(\hat{\gamma}) = P_{\pm}(\hat{\gamma})W(\hat{\gamma})$ and construct
the multiple coefficients using (\ref{multipoledecomposition}) and (\ref{eblm}) (with tilde placed on all the relevant
quantities). It can be verified that the two definitions are equivalent. 

Before proceeding, let us point out some simplifying relations. Firstly, from the definitions of 
$\bar{\eth}_s$ and ${\eth}_s$, it follows that $\bar{\eth}_2=({\eth}_2)^*$ and $\bar{\eth}_1=({\eth}_1)^*$.
In light of definitions (\ref{epsilon}), (\ref{beta}), (\ref{pseudo-epsilon}) and (\ref{pseudo-beta}), it follows
that the two sets of fields  (${\mathcal{E}}$, ${\mathcal{B}}$) and (${\tilde{\mathcal{E}}}$, $\tilde{\mathcal{B}}$)
are real. Furthermore, from the definitions one has
\bea 
\bar{\eth}_1\bar{\eth}_2 P_+ = -{\mathcal{E}}-i{\mathcal{B}},
~~~~
 \bar{\eth}_1\bar{\eth}_2 (P_+W) = -\tilde{{\mathcal{E}}}-i\tilde{{\mathcal{B}}}.
\label{ab1} 
\ena 
Thus, in order to determine the relation between the two sets of fields
(${\mathcal{E}}$, ${\mathcal{B}}$) and ($\tilde{{\mathcal{E}}}$, $\tilde{{\mathcal{B}}}$),
it suffices to study the relation between $\bar{\eth}_1\bar{\eth}_2 P_+$
and $\bar{\eth}_1\bar{\eth}_2 (P_+W)$.

In order to derive the relation between the two sets, it is convenient to expand the
quantity $[\bar{\eth}_1\bar{\eth}_2(P_+W)]W$, using the definition of $\bar{\eth}_s$
operator (\ref{eth1}) in the following way
\bea
[\bar{\eth}_1\bar{\eth}_2(P_+W)]W
&=&(\bar{\eth}_1\bar{\eth}_2P_+)W^2+(\bar{\eth}_1W)(\bar{\eth}_2P_+)W+(\bar{\eth}_2W)(\bar{\eth}_1P_+)W 
\nonumber \\
&&~~+P_+W(\bar{\eth}_1\bar{\eth}_2W)+\cot\theta [W(\bar{\eth}_2P_+)+P_+(\bar{\eth}_2W)]W\nonumber \\
&&~~+(2+2\cot^2\theta)P_+W^2+2\cot\theta W\bar{\eth}_1(P_+W). \ena
Using the following set of relations that follow from (\ref{eth1})
\bea
(\bar{\eth}_2 P_+)W=\bar{\eth}_2 (P_+W)-P_+\bar{\eth}_2 W-2\cot\theta P_+ W, \nonumber \\
(\bar{\eth}_1 P_+)W=\bar{\eth}_1 (P_+W)-P_+\bar{\eth}_1 W-\cot\theta P_+ W, \nonumber \\
W(\bar{\eth}_2 P_+) +P_+(\bar{\eth}_2 W)= \bar{\eth}_2 (P_+
W)-2\cot\theta P_+ W, \nonumber 
\ena 
along with expressions in (\ref{ab1}) we arrive at the expression
\bea
[{\mathcal{E}}+i{\mathcal{B}}] W^2=[\tilde{\mathcal{E}} + i\tilde{\mathcal{B}}] W&+& 
\left\{ \frac{}{} (\bar{\eth}_1 W) [\bar{\eth}_2 (P_+ W)-P_+\bar{\eth}_2 W-2\cot\theta P_+ W] \right.
\nonumber \\
&&~~+(\bar{\eth}_2 W) [\bar{\eth}_1 (P_+ W)-P_+\bar{\eth}_1 W-\cot\theta P_+ W] + P_+ W(\bar{\eth}_1\bar{\eth}_2 W) 
\nonumber \\
&&\left. \frac{}{}~+\cot\theta \bar{\eth}_2(P_+ W)+2W^2 P_+ +2W\cot\theta
\bar{\eth}_1(P_+W)\right\}. \nonumber \\
\ena 
This expression can be rewritten in a compact form
\bea
[{\mathcal{E}}+i{\mathcal{B}}]W^2=[\tilde{{\mathcal{E}}}+i\tilde{{\mathcal{B}}}]W+{\rm ct},
\label{main1} 
\ena 
where ${\rm ct}$ denotes the correction term. This correction term is complex in general.
The real and imaginary parts of the correction term are given as
\bea
{\rm Re}[{\rm ct}]&=&Q[3\cot\theta W W_{x}+W(W_{xx}-W_{yy})-2(W_x^2-W_y^2)]\nonumber\\
&+&U[2\cot\theta W W_y+2WW_{xy}-4W_xW_y]\nonumber\\
&+&2W_x[(QW)_x+(UW)_y]+2W_y[(UW)_x-(QW)_y], 
\label{Re-cts}
\ena 
and
\bea
{\rm Im}[{\rm ct}]&=&U[3\cot\theta W W_{x}+W(W_{xx}-W_{yy})-2(W_x^2-W_y^2)]\nonumber\\
&-&Q[2\cot\theta W W_y+2WW_{xy}-4W_xW_y]\nonumber\\
&-&2W_y[(QW)_x+(UW)_y]+2W_x[(UW)_x-(QW)_y]. 
\label{Im-cts}
\ena
In the above expressions we have introduced the shorthand notations
$F_x\equiv \frac{\partial F}{\partial \theta}$, $F_y\equiv \frac{\partial F}{\sin\theta\partial \phi}$,
$F_{xx}\equiv \frac{\partial^2 F}{\partial \theta^2}$,
$F_{yy}\equiv \frac{\partial^2 F}{\sin^2\theta\partial \phi^2}$
and $F_{xy}\equiv \frac{\partial^2 F}{\sin\theta\partial
\phi\partial\theta}$ for an arbitrary function $F(\hat{\gamma})$. In
Appendix \ref{appendixA}, we discuss the question of numerically calculating
the various terms in the above expression in pixel space.

Finally, one can construct the pure electric and magnetic fields 
${\mathcal{E}}$ and ${\mathcal{B}}$ on the observed portion of the sky (i.e.~region
of the sky for which $W(\hat{\gamma})\neq 0$) using expression
\bea
[{\mathcal{E}}+i{\mathcal{B}}]=[\tilde{{\mathcal{E}}}+i\tilde{{\mathcal{B}}}]W^{-1}+{\rm ct}W^{-2}.
\label{main2} 
\ena 
The construction of the pure electric and magnetic  scalar fields ${\mathcal{E}}$ and ${\mathcal{B}}$ is the main
result of this paper. It is worth pointing out that the construction of these fields is independent of the choice of the
mask function $W(\hat{\gamma})$, as long as the mask in non-zero in the observed portion of the sky. This method
for recovering the scalar fields ${\mathcal{E}}$ and ${\mathcal{B}}$ is lossless in the real space in following sense. 
If one was given the polarization fields $Q$ and $U$ on the entire sphere and constructed the corresponding
$\mathcal{B}$ field using (\ref{beta}) (or the $\mathcal{E}$ field using (\ref{epsilon}) ) and compared the resulting scalar field in the observed region 
with result of above procedure (\ref{main2}), one would find the two fields equal.
However, due to the ill-behaved nature of $W^{-1}$ and $W^{-2}$ at the edge of observed
region, it is difficult to realize the above construction in practice. In order to circumvent this problem, as will be 
discussed in the following section, one has to remove the edge of the constructed polarization maps.

In conclusion of this section it is instructive to clarify the issues associated with possible leakage from the so-called
ambiguous modes. It is known that, on a manifold with a boundary, the decomposition 
of the polarization field, in addition to pure $E$ and $B$ components, contains ambiguous modes that satisfy
both $E$-mode and $B$-mode conditions simultaneously (see \cite{lewis,pixel} for details). In particular, when constructing
the power spectrum estimators for $B$ mode one has to ensure that there is no leakage from the ambiguous 
modes. In the current work, the $\mathcal{B}_{\rm rec}(\hat{\gamma})$ field does not contain contribution from
either $E$ modes of polarization or ambiguous modes, by the virtue of construction 
(analogous to $\chi_{B}$ in \cite{smith2}). For this reason, the power
spectral estimators constructed from this field will be free from contaminations from both $E$-modes and
ambiguous modes.


\section{${\mathcal{E}}/{\mathcal{B}}$ separation in pixel space \label{section3}}

In this section, we shall discuss the issues related to separation of electric and
magnetic polarizations ${\mathcal{E}}(\hat{\gamma})$ and 
${\mathcal{B}}(\hat{\gamma})$ in the pixel space
using the results of the previous section, in particular expression (\ref{main2}). 
We shall discuss this procedure using a toy model. For this toy model, we assume that an
experiment will only observe the Stokes parameters $Q$ and $U$ in the
northern hemisphere. Following \cite{challinor,smith}, we adopt the
following axially-symmetric form for the mask window function $W(\hat{\gamma})$
 \bea
 W(\hat{\gamma})=\left\{
 {\begin{array}{l}
 1~~~~~~~~~~~~~~~~~~~~~~~~~~~~~~~~~~~~\theta<\theta_0-\theta_1,  \\
 \frac{1}{2}-\frac{1}{2}\cos(\frac{\theta-\theta_0}{\theta_1}\pi)
 ~~~~~~~~~~~~~~~~\theta_0-\theta_1<\theta<\theta_0,  \label{window} \\
 0~~~~~~~~~~~~~~~~~~~~~~~~~~~~~~~~~~~~\theta>\theta_0.
 \end{array}
 }
 \right.
 \label{windowfunction}
\ena
In the above expression, $\theta_0$ corresponds to edge of the observational area, and
$\theta_1\geq 0$ is the smoothing scale. The limiting case, $\theta_1=0$, corresponds to the top-hat window function
($w(\hat{\gamma})=1$ for $\theta<\theta_0$ and $w(\hat{\gamma})=0$ for $\theta>\theta_0$).
However, the top-hat function is discontinuous at $\theta=\theta_0$, which makes quantities
$W_{xx}$, $W_{xy}$ and $W_{yy}$ ill-defined at the edge. In order to avoid these difficulties,
it is convenient to use a window function with $\theta_1\neq0$. Throughout the present section
we use values $\theta_0=90^o$ and $\theta_1=30^o$. It is important to point out that the formalism
outlined in Sec.~\ref{section2} is applicable for arbitrary mask window functions, not necessarily axially symmetric.
The simple symmetric form (\ref{windowfunction}) for the mask was chosen for simplicity and clarity of presentation.
In realistic scenarios one will have to use a more complex mask, that will take into account the non-symmetric form of 
the observed region and various point source contaminations.

For an axially symmetric window function $W(\hat{\gamma})$ (i.e.~when $W(\hat{\gamma})$ is independent of $\phi$),
such as the one considered in the present section, the correction term ${\rm ct}$ in (\ref{Re-cts}) and (\ref{Im-cts})
simplify to
\bea
{\rm Re}[\rm{ct}]&=&Q[3\cot\theta W W_{x}+WW_{xx}-2W_x^2]+2W_x[(QW)_x+(UW)_y], \label{real-cts}\\
{\rm Im}[\rm{ct}]&=&U[3\cot\theta W W_{x}+WW_{xx}-2W_x^2]
+2W_x[(UW)_x-(QW)_y]. 
\label{imagine-cts}
\ena

In order to demonstrate the ${\mathcal{E}}/{\mathcal{B}}$ separation we shall work with simulated polarization maps.
We use the {\it synfast} subroutine included in the HEALPix package to generate a full sky map of 
$Q(\hat{\gamma})$ and $U(\hat{\gamma})$ fields. In order to generate this map we use the best-fit WMAP5 
values for cosmological parameters \cite{wmap5}
\bea
\Omega_b h^2=0.02267,~\Omega_c
 h^2=0.1131,~\Omega_{\Lambda}=0.726,~\nonumber\\
 \tau_{\rm reion}=0.084,~A_s=2.446\times10^{-9},~n_s=0.96,
\label{background}
\ena
and assume no contribution from gravitational waves and cosmic lensing 
to the $B$-mode of polarization, i.e.~$C_{\ell}^{BB}=0$. We adopt the pixelization with
$N_{\rm side}=512$. We set the full width at half maximum (FWHM) for the Gaussian beam to $\theta_F=30'$.
In what follows we shall be solely interested in the determination of $\mathcal{B}$ polarization field,
and studying the possible contaminations to it. In this context, since $C_{\ell}^{BB}=0$, 
a non-zero field $\mathcal{B}$ would be wholly attributed to the residual $EB$-mixing contamination.

\begin{figure}[t]
\centerline{\includegraphics[width=8cm]{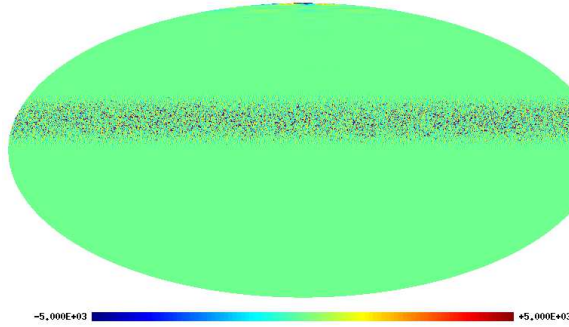}} \caption{The
$\tilde{\mathcal{B}}$ map constructed from an input model with no
magnetic polarization (in $\mu$K).} \label{fig1}
\end{figure}

Before proceeding to construct the pure magnetic field $\mathcal{B}$, we shall construct and analyze the auxiliary $\tilde{{\mathcal{B}}}$ field.
From the simulated $Q(\hat{\gamma})$ and $U(\hat{\gamma})$ maps we construct the 
$\tilde{{\mathcal{B}}}(\hat{\gamma})$ in the following manner. Firstly, we construct 
the multipole coefficients $\tilde{B}_{\ell m}$. This is done by building multipole coefficients
$\tilde{a}_{\pm2,\ell m} = \int d \hat{\gamma}~P_{\pm}(\hat{\gamma})W(\hat{\gamma})~ _{\pm2}Y^*_{\ell m}
(\hat{\gamma}) $ using the simulated ($Q(\hat{\gamma})$, $U(\hat{\gamma})$) maps 
along with the window function $W(\hat{\gamma})$ given in (\ref{windowfunction}), and then calculating
$ \tilde{B}_{\ell m} = -\frac{1}{2i}\left[\tilde{a}_{2,\ell m}-\tilde{a}_{-2,\ell m}\right]$. These steps were performed numerically
using the {\it anafast} routine from the HEALPix package. Following this,
we use (\ref{pseudo-epsilonbeta}) to construct $\tilde{\mathcal{B}}$ from the multipole coefficients 
$\tilde{B}_{\ell m}$. The resulting $\tilde{\mathcal{B}}$ map is illustrated in Fig.~\ref{fig1}.

One can see that although the map was generated with $C_{\ell}^{BB}=0$, $\tilde{\mathcal{B}}\neq0$
in the region $\theta_0-\theta_1<\theta<\theta_0$ (i.e. $60^o<\theta<90^o$). This can be viewed as a 
of leakage of $\mathcal{E}$-type of polarization into $\tilde{\mathcal{B}}$ field, due to the presence
of the window function $W$. In order to quantify this leakage in the harmonic (multipole) space, we construct the pseudo power spectrum
as
\bea
D_{\ell}^{\tilde{{\mathcal{B}}}\tilde{{\mathcal{B}}}}\equiv\frac{1}{2\ell+1}\sum_{m}
\tilde{{\mathcal{B}}}_{\ell m}\tilde{{\mathcal{B}}}^*_{\ell m},~~~ \rm{where}~
\tilde{{\mathcal{B}}}_{\ell m}\equiv\int \tilde{{\mathcal{B}}}(\hat{\gamma})Y^*_{\ell
m}(\hat{\gamma})d\hat{\gamma}.
\label{pseudoresidualestimators}
\ena
The resulting pseudo power spectrum is plotted a the black line in Fig.~\ref{fig3}.

\begin{figure}[t]
\centerline{\includegraphics[width=8cm]{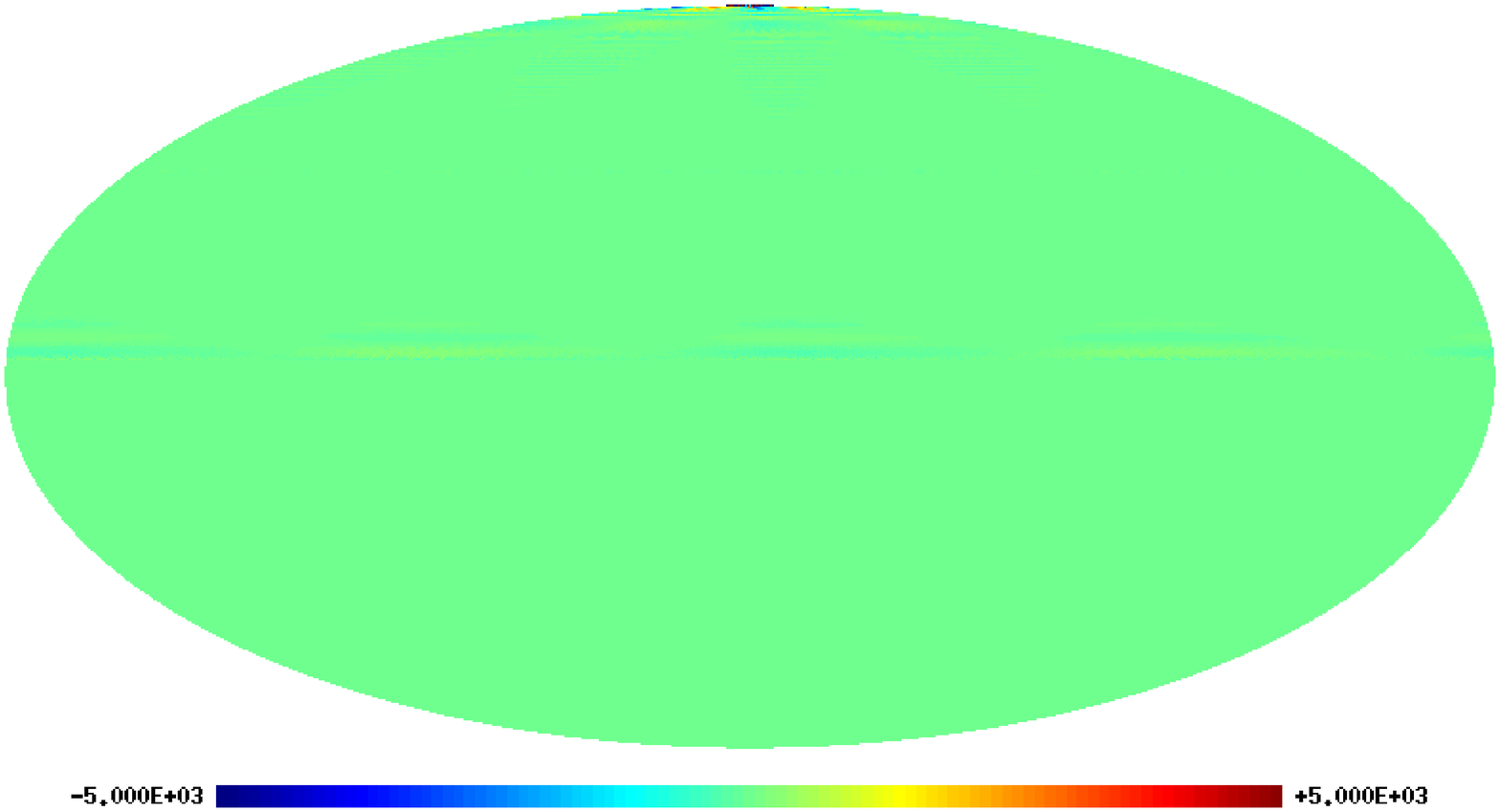}\includegraphics[width=8cm]{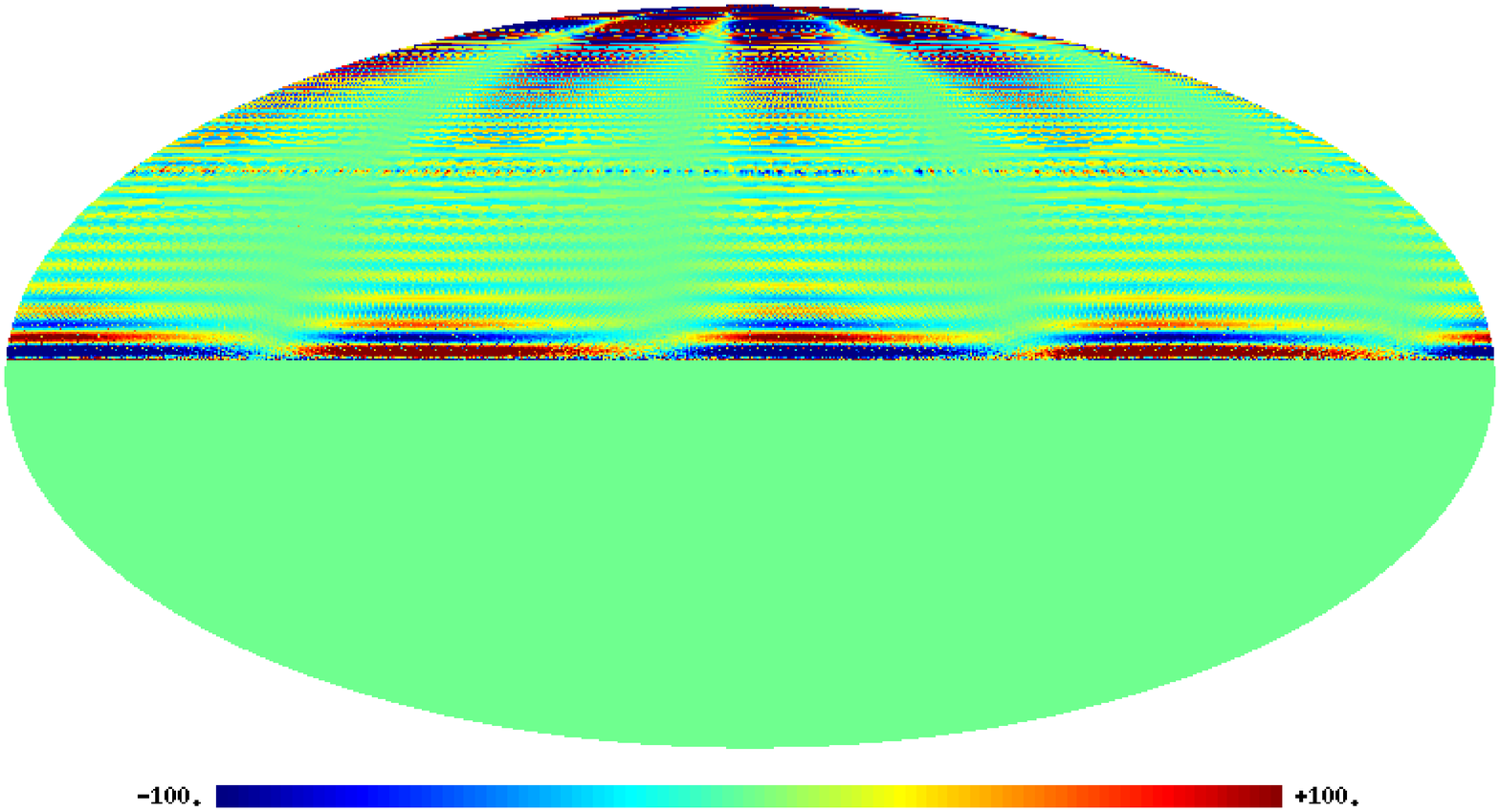}}
\caption{The pure magnetic field ${\mathcal{B}}_{\rm rec}(\hat{\gamma})$. The edge of the map
with $W(\hat{\gamma})<0.03$ has been removed. The left panel is scaled similarly to Fig.~\ref{fig1},
while the right panel has the scaling magnified in order to show the residual leakage.
Both the panels use the units $\mu$K.} 
\label{fig2}
\end{figure}

\begin{figure}[t]
\centerline{\includegraphics[width=14cm]{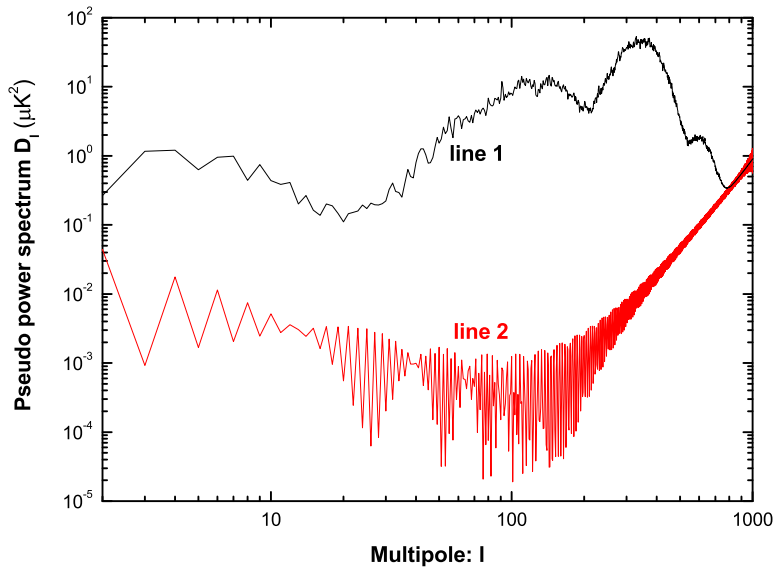}} \caption{The pseudo
power spectrum $D_{\ell}^{\tilde{{\mathcal{B}}}\tilde{{\mathcal{B}}}}$ (black line, i.e. line 1) of the $\tilde{\mathcal{B}}$ field
and pseudo power spectrum $D_{\ell}$ (red line, i.e. line 2) of the ${\mathcal{B}}_{\rm rec}$ field.}
\label{fig3}
\end{figure}

We can now reconstruct pure magnetic type field $\mathcal{B}$ using (\ref{main2}) in the following
manner 
\bea
\mathcal{B}_{\rm rec}(\hat{\gamma})\equiv \tilde{\mathcal{B}}(\hat{\gamma})
W^{-1}(\hat{\gamma})+{\rm Im}({\rm ct}(\hat{\gamma}))W^{-2}(\hat{\gamma}). 
\label{main3}
\ena
We use the subscript ${\rm rec}$ to indicate that this field was reconstructed 
field from $\tilde{\mathcal{B}}$ and the ($Q$, $U$) fields. The results of reconstruction are presented
in Fig.~\ref{fig2}. Since the input cosmological
model assumes no contribution from gravitational waves, one can expect $\mathcal{B}_{\rm rec}(\hat{\gamma})=0$.
A visual comparison of Fig.~\ref{fig1} and Fig.~\ref{fig2} shows that we have been able to remove
much of the leakage that was present in $\tilde{\mathcal{B}}$. The remaining residual contamination in $\mathcal{B}_{\rm rec}$
is shown (with a magnified scale) in the right panel of Fig.~\ref{fig2}. This residuals are a small fraction ($\sim 2\%$) of 
the total leakage in Fig.~\ref{fig1}. In order to quantify these residuals in multipole space, we construct the pseudo spectral estimators
replacing $\tilde{\mathcal{B}}$ with $\mathcal{B}_{\rm rec}$ in (\ref{pseudoresidualestimators}). The resulting 
pseudo power spectrum is plotted as the red curve in Fig.~\ref{fig3}. It can be seen that the resulting leakage
power for $\mathcal{B}_{\rm rec}$ is significantly lower that the corresponding power for $\tilde{\mathcal{B}}$.
In particular, in the practically interesting range of multipoles $\ell\in(50,~200)$, the spectrum for the reconstructed
 $\mathcal{B}_{\rm rec}$ field is roughly four order of magnitude lower than the spectrum for $\tilde{\mathcal{B}}$.
 The remaining residuals in $\mathcal{B}_{\rm rec}$ are attributed to numerical errors.

We believe that the small remaining residuals in $\mathcal{B}_{\rm rec}$ are a result of two types of numerical errors
that cannot be avoided in practice. The first reason for errors is pixelization. In \cite{pixel}, it was argued that pixelization can
lead to the mixing of electric and magnetic modes. This point can be intuitively understood in the following way.
Imagine a survey that observes polarization on a small square region of the sky. Pixelization introduces a Nyquist wavenumber
$k_{N}$, such that all modes with wavenumbers greater than $k_N$ are aliased to modes with wavenumbers less than
the Nyquist value. This aliasing completely shuffles the direction of wavenumbers, thus essentially leading to a complete
mixing of electric and magnetic modes. Although the complete avoidance of the errors due to pixelization is impossible,
these numerical errors can be reduced by using a larger value for $N_{\rm side}$  (see Sec.~\ref{section3.2} for details). In
the present evaluation, with $N_{\rm side}=512$, finite pixelization seems to be the main reason for residual power spectrum for $\ell>150$. 
The main contribution to this residual power spectrum comes from the residual $\mathcal{B}_{\rm rec}(\hat{\gamma})$ 
at the pole in the real space (see the right panel in Figure \ref{fig2}).

The second reason for numerical errors is the steep growth of power spectrum $C_{\ell}^{{\mathcal{E}}{\mathcal{E}}}$ with
increasing $\ell$. Due to this, even a small relative numerical error at higher multipoles seeps through to lower multipoles.
In other words, these errors occur due to fact that the various sources of noise and $E$ mode signal are not band limited.
We believe that these types of errors mainly account for the residual power spectrum of $\mathcal{B}_{\rm rec}$
at multipoles $\ell<150$. These errors are mainly caused by the residual $\mathcal{B}_{\rm rec}(\hat{\gamma})$ 
around the observed edge $\theta=\theta_0$ in the real space (see the right panel in Figure \ref{fig2}).

\subsection{Dependence of residual leakage on edge removal \label{section3.1}}

Figure~\ref{fig2} shows that much of the residual leakage occurs around the edge of the observation region. 
It is therefore instructive to study the edge effects in more detail.
The expression (\ref{main3}) for $\mathcal{B}_{\rm rec}$ depends on
${\rm Im({\rm ct})}$ correction term given by (\ref{imagine-cts}). This correction term contains
two terms, $(3\cot\theta W W_{x}+WW_{xx}-2W^2_{x})/W^2$ and $2W_{x}/W^2$, which have
the following asymptotic at the edge of the map as $\theta\rightarrow \theta_0$
\bea 
\frac{3\cot\theta W W_{x}+WW_{xx}-2W^2_{x}}{W^2}\rightarrow
-\frac{6}{(\theta_0-\theta)^2}, 
~~~~
\frac{2W_{x}}{W^2}\rightarrow
-\frac{16\theta_1^2}{\pi^2(\theta_0-\theta)^3}. 
\nonumber
\ena 
Thus, the two functions are divergent for $\theta\rightarrow \theta_0$. This implies that the signal-to-noise
will tend to zero for data as the boundary of the observed region is approached.
Because of this divergence, in numerical calculations, one must remove the edge of the map in order 
to avoid numerical errors associated with these divergences, thereby introducing a small loss in information.

\begin{figure}[t]
\centerline{\includegraphics[width=14cm]{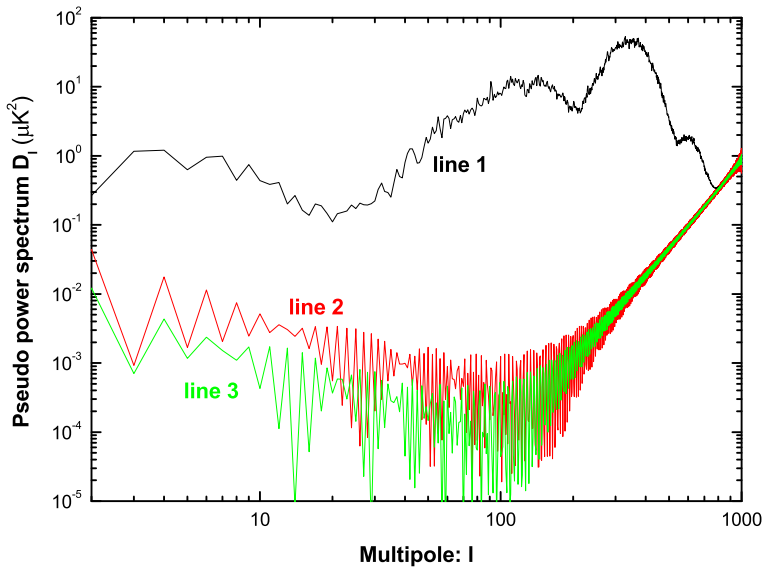}} \caption{The
red line (i.e. line 2) shows the pseudo power spectrum $D_{\ell}$ calculated for
the edge removal at $W=0.03$, while the green line (i.e. line 3) shows the
$D_{\ell}$ for edge removal at $W=0.1$. For
comparison, the black line (i.e. line 1) shows the pseudo power spectrum
$D_{\ell}^{\tilde{{\mathcal{B}}}\tilde{{\mathcal{B}}}}$. Note that, the black line (i.e. line 1) and red line (i.e. line 2) 
are identical to those in Fig.~\ref{fig3}.} 
\label{fig4}
\end{figure}

In order to investigate the dependence of residual leakage on the edge removal, in Fig.~\ref{fig4}
we plot the pseudo power spectrum $D_{\ell}$ of residual $\mathcal{B}_{\rm rec}$ constructed for
the same simulated data but for two different edge removals. The first case (red line) corresponds
to portion of the sky with $W<0.03$ removed (this corresponds to removal of data with $\theta_0-\theta<0.06$).
The second case (green line) corresponds to a portion of sky with $W<0.1$ removed 
(corresponding to removal of data with $\theta_0-\theta<0.1$). The second case corresponds to 
a larger portion of the sky removed, and thus to a larger loss in information. This loss of information
leads to a smaller value for the power spectrum at lower multipoles $\ell<150$. This fact can 
be clearly seen from right panel of Fig.~\ref{fig2}. As one removes more of the data from the equator
the residual spectrum of $\mathcal{B}_{\rm rec}$ becomes smaller. On the other hand,
in the region of higher multipoles, where the dominant contribution comes from finite pixelization errors,
the two power spectra are comparable.

\subsection{Dependence of residual leakage on pixelization number $N_{\rm side}$ \label{section3.2}}

\begin{figure}[t]
\centerline{\includegraphics[width=14cm]{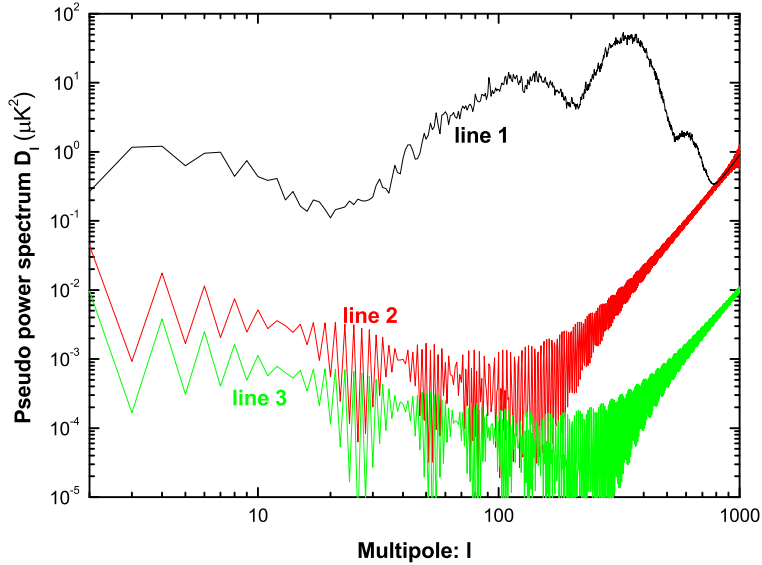}} \caption{The
red line (i.e. line 2) shows the pseudo power spectrum $D_{\ell}$ calculated for $N_{\rm side}=512$, 
while the green line (i.e. line 3) shows $D_{\ell}$ calculated for $N_{\rm side}=1024$.
For comparison, the black line (line 1) shows the pseudo power spectrum
$D_{\ell}^{\tilde{{\mathcal{B}}}\tilde{{\mathcal{B}}}}$. Note that, the black line (i.e. line 1) and red line (i.e. line 2) 
are identical to those in Fig.~\ref{fig3}.} 
\label{fig5}
\end{figure}

As was pointed out earlier, one of the reasons for residual leakage of power into $\mathcal{B}_{\rm rec}$
is finite pixelization of the sky map. In order to demonstrate the effect of pixelization on the residual
power, in Fig.~\ref{fig5} we show the pseudo power spectrum $D_{\ell}$ calculated for two different
pixelization numbers $N_{\rm side}=512$ (red line) and $N_{\rm side}=1024$ (green line). As one might expect,
the increase in the pixelization number reduces the leakage power spectrum. This reduction is most
dramatic at higher multipoles $\ell>150$, where it is two orders of magnitude for this example. At lower
multipoles $\ell<150$ the reduction is not as dramatic, and is roughly by a factor $3$. These results are
consistent with our previous statements about the cause of numerical errors. Indeed, at higher multipoles
the main cause of errors seems to be finite pixelization, whereas at lower multipoles the errors are generated
by a combination of factors.

\subsection{Dependence of residual leakage on $\theta_{F}$ \label{section3.3}}

The full width at half maximum parameter $\theta_F$ has an important effect on the residual
leakage of power into $\mathcal{B}_{\rm rec}$. In order to understand the reason for this,
one has to remember that one of the two reasons for residual leakage is the steep growth
of power spectrum $C_{\ell}^{{\mathcal{E}}{\mathcal{E}}}$ with increasing $\ell$. The
parameter  $\theta_F$ regulates the exponential damping of this power spectrum at 
multipoles $\ell\simeq\theta_F^{-1}$, and therefore limits the propagation of the power
in $C_{\ell}^{{\mathcal{E}}{\mathcal{E}}}$ into $D_{\ell}$.

\begin{figure}[t]
\centerline{\includegraphics[width=14cm]{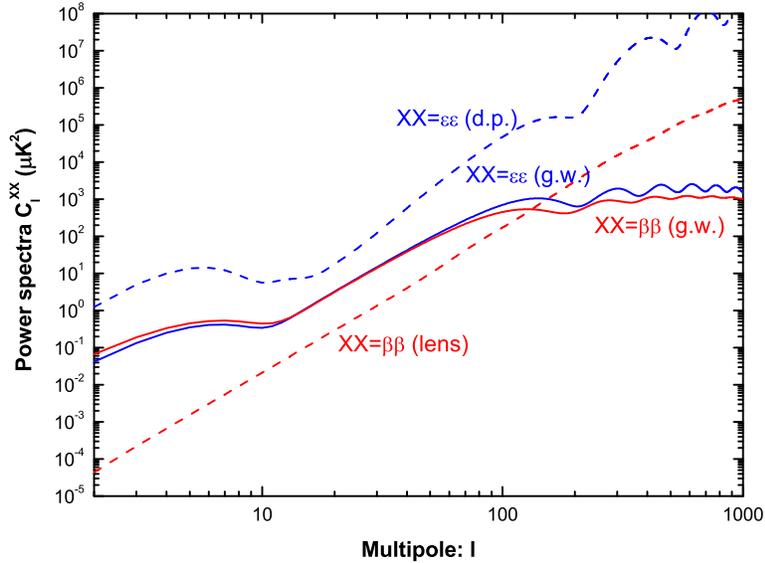}}
\caption{The polarization power spectra generated by density
perturbations (d.p.), gravitational waves (g.w.) with $r=0.1$,
and cosmic lensing (lens).} 
\label{fig6}
\end{figure}

The various  contributions to the spectrum $C_{\ell}^{{\mathcal{E}}{\mathcal{E}}}$ 
are illustrated in Fig.~\ref{fig6}. The main contribution to power
spectrum comes from density perturbations (dashed blue line). For comparison, on this figure, we show the contribution
to $C_{\ell}^{{\mathcal{E}}{\mathcal{E}}}$ (solid blue line) and $C_{\ell}^{{\mathcal{B}}{\mathcal{B}}}$ (solid red line) from
gravitational waves (characterized by tensor-to-scalar ratio $r=0.1$), as well as contribution to $C_{\ell}^{{\mathcal{B}}{\mathcal{B}}}$ 
from lensing (red dashed line). The spectrum $C_{\ell}^{{\mathcal{E}}{\mathcal{E}}}$ from density perturbations at high multipoles
acts as the main source for the residual leakage into $D_{\ell}$.

\begin{figure}[t]
\centerline{\includegraphics[width=14cm]{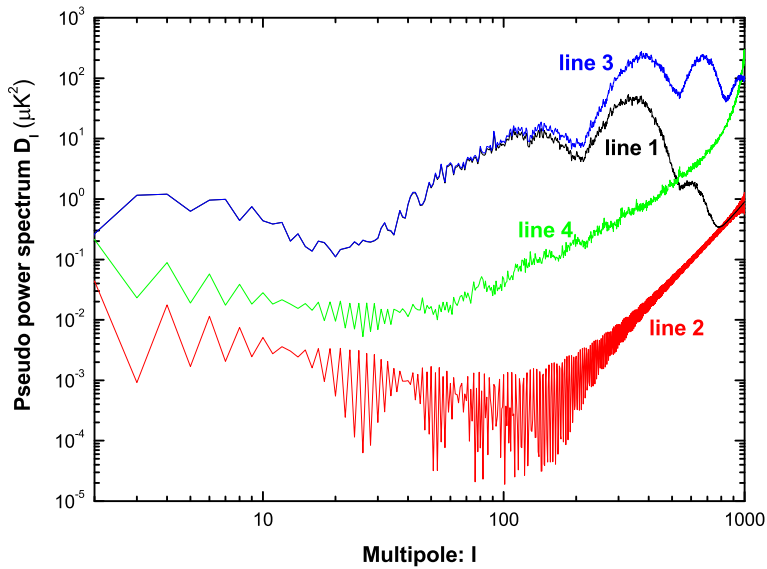}} 
\caption{
The red line (i.e. line 2) shows the pseudo power spectrum $D_{\ell}$, calculated for
a map with $\theta_F=30'$,  while the green line (i.e. line 4) shows $D_{\ell}$ calculated
with $\theta_F=10'$. The black line (i.e. line 1) shows $D_{\ell}^{\tilde{{\mathcal{B}}}\tilde{{\mathcal{B}}}}$
calculated for $\theta_F=30'$, and the blue line (i.e. line 3) shows $D_{\ell}^{\tilde{{\mathcal{B}}}\tilde{{\mathcal{B}}}}$
for $\theta_F=10'$. Note that, the black line (i.e. line 1) and red line (i.e. line 2) 
are identical to those in Fig.~\ref{fig3}.}
\label{fig7}
\end{figure}

The contribution to the various spectra at high multipoles are effectively damped by a choice of an appropriate $\theta_F$.
This parameter leads to the damping of power spectrum $C_{\ell}^{{\mathcal{E}}{\mathcal{E}}}$ proportional to  
$\exp\left(-\frac{\ell(\ell+1)\theta_F^2}{8\ln 2}\right)$. In Fig.~\ref{fig7} we show the residual leakage for two choices of the FWHM parameter
$\theta_F=30'$ (red line) and $\theta_F=10'$ (green line). For comparison, in this figure we also show the pseudo spectrum
$C_{\ell}^{\tilde{\mathcal{B}}\tilde{\mathcal{B}}}$ calculated $\theta_F=30'$ (black line) and $\theta_F=10'$ (blue line). 
As one might expect, the residual power spectrum reduces significantly with an increase in $\theta_F$. For this reason, for the purposes
of extracting the magnetic pattern of polarization in experiments with small $\theta_F$ (for example POLARBEAR experiment discussed in
Sec.~\ref{section4.4}), it becomes necessary to artificially increase $\theta_F$ in order to reduce residual leakage in ${\mathcal{B}}_{\rm rec}$.
In Appendix~\ref{appendixB}, we suggest a `map smoothing' technique to achieve this goal.

\section{${\mathcal{E}}/{\mathcal{B}}$ separation and power spectrum estimation for small sky surveys \label{section4}}

In Sec.~\ref{section2} and Sec.~\ref{section3} we developed a method to construct 
pure electric $\mathcal{E}_{\rm rec}$ and magnetic fields $\mathcal{B}_{\rm rec}$
out of the original Stokes parameter fields $Q$ and $U$ on a fractional portion of the sky. Ignoring the small numerical
errors, it was shown that the resulting fields did not exhibit mixing. A crucial point about the constructed fields
is that they are scalar fields. For this reason, one can use all of the robust techniques developed for studying
CMB temperature anisotropy to the fields $\mathcal{E}_{\rm rec}$ and $\mathcal{B}_{\rm rec}$. 
Based on appropriation of these techniques, in this section, we shall focus on an important practical application,
namely constructing the estimator for the power spectrum of the $B$-mode of polarization $C_{\ell}^{BB}$. 
For this reason, as in the previous sections, we shall restrict our analysis to just the magnetic field $\mathcal{B}_{\rm rec}$.

The question of constructing an estimator for the power spectrum $C_{\ell}^{BB}$ from the field $\mathcal{B}_{\rm rec}$ is analogous
to the problem of construction an estimator for the temperature anisotropy power spectrum $C_{\ell}^{TT}$ given a temperature map
on a partial sky. Fortunately, there are a large number of methods that have been developed for this purpose \cite{tegmark,tegmarkPolarization,NRML,old-pseudo,hybrid}.
Amongst these, a popular method is the so-called `pseudo-$C_{\ell}$' estimator method \cite{old-pseudo}. This method can be easily 
realized in pixel space using fast spherical harmonics transformation, and has been applied to various CMB observations including WMAP data \cite{wmap-pseudo}.
However, it is well known that pseudo-$C_{\ell}$ estimators are sub-optimal, particularly for low multipoles. For this reason, many authors have developed
alternative estimators that are optimal, in particular the maximum likelihood estimators in pixel realization \cite{tegmark,NRML}. The fundamental problem
with the maximum likelihood estimators is that these methods are very slow, especially for larger multipoles. For large sky surveys, such as the Planck satellite,
the use of hybrid estimators, which combine the two methods, has been suggested \cite{hybrid}. The hybrid estimator combines the best of two worlds, it is nearly optimal and
can be realized of a laptop computer even for large sky surveys such as {Planck}.

In the present section we shall focus on small sky polarization surveys, corresponding to various ground-based CMB experiments \cite{quad,bicep,clover,POLARBEAR,quiet}.
Since these surveys will be primarily sensitive to relatively large multipoles $\ell\gtrsim 20$, we shall restrict our analysis to pseudo-$C_{\ell}$ type estimators, which are
nearly optimal for large multipoles. Before proceeding we would like to point out that a hybrid type of estimator could be potentially used to construct an estimator for $C_{\ell}^{BB}$
from $\mathcal{B}_{\rm rec}$ in the case of large sky surveys such as Planck. We leave this exercise for future. 

Below we shall work with a small fraction of the sky characterized by a window function (\ref{windowfunction}) with $\theta_0=20^o$ and $\theta_1=10^o$,
corresponding to a $3\%$ sky survey. In an ideal case, neglecting numerical errors, the reconstructed field
$\mathcal{B}_{\rm rec}(\hat{\gamma})$ would be related to the underlying full sky field $\mathcal{B}(\hat{\gamma})$ through 
$\mathcal{B}_{\rm rec}(\hat{\gamma})=\mathcal{B}(\hat{\gamma})w(\hat{\gamma})$, where $w(\hat{\gamma})$ is the corresponding top-hat window function.
However, as was pointed out in Sec.~\ref{section3.1}, one needs to remove a narrow edge from the observational area 
in order to avoid excessive numerical errors. For this reason, in practice, we remove the region $\theta_0-\theta<0.03$ (corresponding to $\theta>18^o$)
from the analysis. Below we shall use the notation $w'(\hat{\gamma})$ to denote the top-hat window function for data with edge removal.


\subsection{Pseudo estimators\label{section4.1}}

The first step in constructing the pseudo estimator is the definition of spherical harmonics coefficients $a_{\ell m}$ of the scalar field ${\mathcal{B}}_{\rm rec}(\hat{\gamma})$ as follows
\bea
a_{\ell m} = \int d\hat{\gamma}~ {\mathcal{B}}_{\rm rec}(\hat{\gamma}) \mathcal{W}(\hat{\gamma}) Y^*_{\ell m}(\hat{\gamma}),
\label{almrec-weighted}
\ena
where $\mathcal{W}(\hat{\gamma})$ is the weight function. In principle, one can choose an arbitrary form for the weight function. In particular, the choice $\mathcal{W}(\hat{\gamma})=1$ corresponds to the widely discussed pseudo-$C_{\ell}$ estimator introduced in \cite{old-pseudo}. This choice will be the main focus of our attention in the present work. An alternative choice $\mathcal{W}(\hat{\gamma})=W(\hat{\gamma})$ (where $W(\hat{\gamma})$ is the mask window function in Eq. (\ref{windowfunction}))
corresponds to the analysis in \cite{smith2}, where it was shown that the resulting $a_{\ell m}$ lead to the pure $B$-mode estimators defined in \cite{smith}. The comparison of this choice for the weight function with our main choice $\mathcal{W}(\hat{\gamma})=1$ is discussed in Appendix \ref{appendixA3}. The optimal choice of the weight function in various cases has been discussed in \cite{challinor,smith2,mask3}. In \cite{smith2} the authors suggest a general method to build the weight function $\mathcal{W}(\hat{\gamma})$ for different multipole $\ell$ in order to optimize the estimator. At this point it is important to emphasize that although ${\mathcal{B}}_{\rm rec}(\hat{\gamma})$ preserves the available information in real space, a non-optimal power spectrum estimation will lead to loss of some of this information. This makes the study of the optimal choice of weight function particularly important. However, in the current paper we concentrate mainly on the simplistic case $\mathcal{W}(\hat{\gamma})=1$, leaving the important but complicated question of optimal choice of weight function for future work.

For the choice $\mathcal{W}(\hat{\gamma})=1$, the spherical harmonics coefficients $a_{\ell m}$ in (\ref{almrec-weighted}) take the simplified form
\bea
a_{\ell m} = \int d\hat{\gamma}~ {\mathcal{B}}_{\rm rec}(\hat{\gamma}) Y^*_{\ell m}(\hat{\gamma}).
\label{almrec}
\ena
These are related to the coefficients ${\mathcal{B}}_{\ell m}$ (which were defined in (\ref{mathcalelmblm}) in terms of the underlying
full sky map ${\mathcal{B}}(\hat{\gamma})$) through the coupling matrix $K_{\ell m \ell' m'}$ (see for instant \cite{hybrid})
\bea
 {a}_{\ell m} = \sum_{\ell' m'} B_{\ell'}{\mathcal{B}}_{\ell'
 m'} K_{\ell m \ell' m'} = \sum_{\ell' m'} B_{\ell'} N_{\ell'} B_{\ell'
 m'} K_{\ell m \ell' m'}, \label{almblm} 
\ena
where $B_{\ell}$ is a window function describing the combined
smoothing effects of the beam and the finite pixel size. 
The coupling matrix $K$ can be expressed in terms of the
function ${w'}(\hat{\gamma})$ as
\bea
 K_{\ell_1 m_1 \ell_2 m_2}= \int d\hat{\gamma}~{w'}(\hat{\gamma})Y_{\ell_1 m_1}
 (\hat{\gamma}) Y^*_{\ell_2 m_2} (\hat{\gamma}). \label{klmlm}
\ena

The pseudo estimator $D_{\ell}$ is defined analogous to (\ref{pseudoresidualestimators}) in terms of the multipole coefficients (\ref{almrec}) as
\bea
D_{\ell} = \frac{1}{2\ell+1}\sum_{m}a_{\ell m}a^*_{\ell m}.
\label{pseudoDestimator}
\ena
Using relations (\ref{ebpowerspectrum}), (\ref{new-old-relations}) and (\ref{almblm}),
one obtains that the expectation value of this estimator $D_{\ell}$ is
related to the true power spectrum $C_{\ell}^{BB}$ by the following
convolution
\bea
 \langle {D}_{\ell} \rangle = \sum_{\ell'}  M_{\ell \ell'}
 B_{\ell'}^2 C_{\ell'}^{{\mathcal{B}}{\mathcal{B}}} = \sum_{\ell'}  M_{\ell \ell'}
 N^2_{\ell'} B_{\ell'}^2 C_{\ell'}^{BB}.
\label{peudo-estimator}
\ena
The coupling matrix $M$ in the above expression can be expressed in terms of $3j$ symbols as
 \bea
 M_{\ell_1\ell_2} = (2\ell_2+1) \sum_{\ell_3}
 \frac{(2\ell_3+1)}{4\pi} w'_{\ell_3}
 {\left(\begin{array}{ccc}
 \ell_1 & \ell_2 & \ell_3 \\
  0 & 0 & 0
  \end{array}
 \right)^2}, \label{mll}
 \ena
where $w'_{\ell}$ is the power spectrum of the window function  
${w'}(\hat{\gamma})$ defined in an analogous manner to (\ref{pseudoresidualestimators}).

It can be shown that the covariance matrix for the pseudo estimator $D_{\ell}$ has the form
\bea
\langle \Delta{D}_{\ell}\Delta{D}_{\ell'}\rangle =
\frac{2}{(2\ell+1)(2\ell'+1)}\sum_{mm'}\sum_{\ell_1 m_1} \sum_{\ell_2 m_2} && 
B_{\ell_1}^2 N_{\ell_1}^2 C_{\ell_1}^{BB} B_{\ell_2}^2 N_{\ell_2}^2 C_{\ell_2}^{BB} \times \nonumber \\
&& \times K_{\ell m \ell_1 m_1} K^*_{\ell' m' \ell_1 m_1} K^*_{\ell m \ell_2 m_2} K_{\ell' m' \ell_2 m_2}.
\nonumber \\
\label{covariancematrix}
\ena
As it stands, this formula is not useful due to the high cost of
computation. However, for high multipoles, this formula simplifies to \cite{hybrid}
\bea
\langle \Delta{D}_{\ell}\Delta{D}_{\ell'}\rangle \approx {2} B_{\ell}^2 N_{\ell}^2 C_{\ell}^{BB}
B_{\ell'}^2 N_{\ell'}^2 C_{\ell'}^{BB} M_{\ell \ell'} / {(2\ell'+1)}. 
\label{covariance-approximation}
\ena

\begin{figure}[t]
\centerline{\includegraphics[width=14cm]{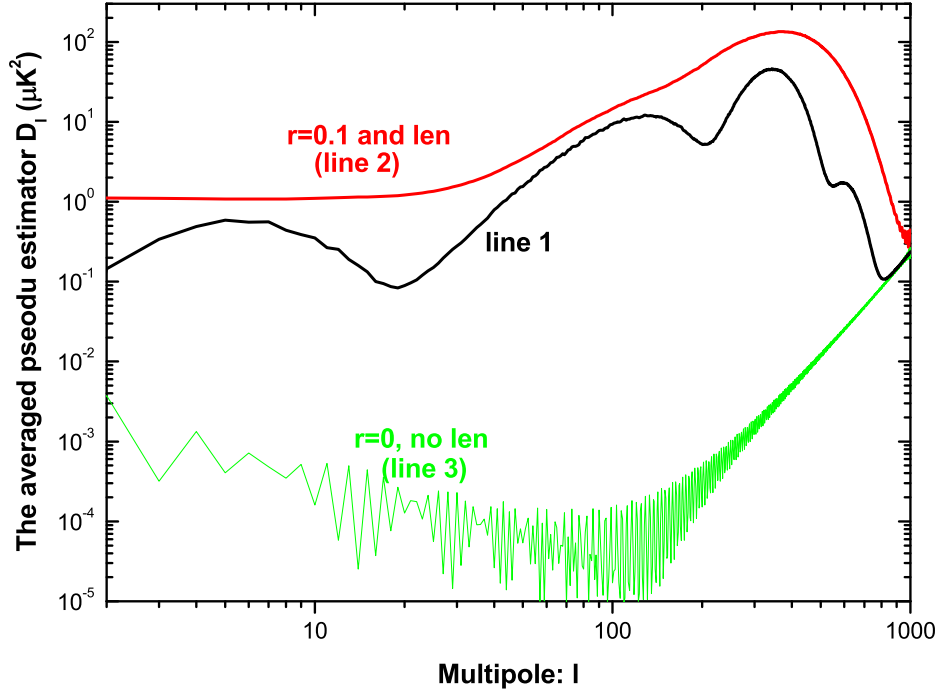}}
\caption{The averaged pseudo estimator $D_{\ell}$ from 1000
simulations. The red line (i.e. line 2) shows the result for an input model
with $r=0.1$ and contribution from cosmic lensing.
The green line (i.e. line 3) shows the result for an input
model with no magnetic polarization (i.e.~$C_{\ell}^{BB}=0$). 
For comparison, the black line (i.e. line 1) shows the averaged estimator
$D_{\ell}^{\tilde{{\mathcal{B}}}\tilde{{\mathcal{B}}}}$ for an input model
with no magnetic polarization.} 
\label{fig8}
\end{figure}

In order to implement and verify the above analytical results we have conducted numerical calculations using simulated data.
In the first instance, we generate 1000 random full sky ($Q$, $U$) maps with no contribution from gravitational waves (i.e.~$r=0$)
and no lensing. For each realization, we reconstruct the magnetic field ${\mathcal{B}}_{\rm rec}(\hat{\gamma})$ and evaluate the pseudo estimator $D_{\ell}$.
The average over 1000 realizations $\overline{D_{\ell}}$ is plotted (green line) in Fig.~\ref{fig8}.
Note that, here and below we use the over-line to denote averaging over simulated realizations, as opposed to the angle
brackets which denote ensemble averaging. The average for the
uncleaned spectrum $\overline{D_{\ell}^{\tilde{{\mathcal{B}}}\tilde{{\mathcal{B}}}}}$ (defined in (\ref{pseudoresidualestimators})) 
is plotted (black line) for comparison on the same figure. For next calculations, we simulate 1000 random full sky maps
with contribution from gravitational waves characterized by $r=0.1$ and contribution to $B$-mode of polarization from cosmic lensing.
The average value of the estimator $\overline{D_{\ell}}$ is plotted (red line) in Fig.~\ref{fig8}. The comparison of curves in Fig.~\ref{fig8} shows
that the residual noise contribution to the pseudo estimator due to numerical errors (green line) 
is negligible in comparison with the contribution to the estimator from the signal (red line). One can therefore conclude that the resulting
pseudo estimator $\overline{D_{\ell}}$ is effectively free from $EB$-mixing.

\begin{figure}[t]
\centerline{\includegraphics[width=14cm]{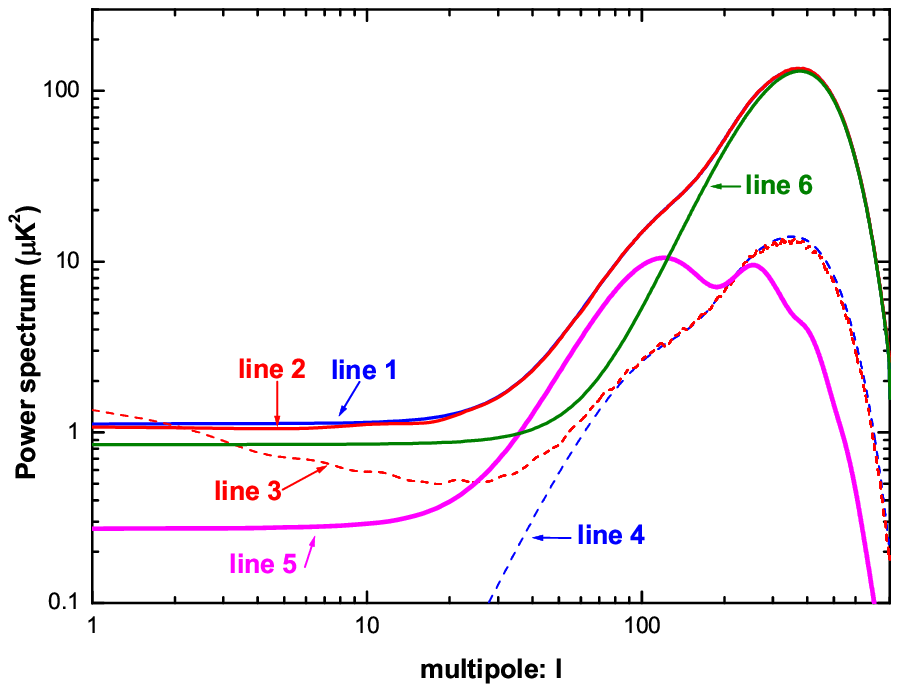}}
\caption{
All of the lines are a result of averaging over 1000 simulated samples.
The red line (i.e. line 2) shows the averaged pseudo estimator $D_{\ell}$ for an input model 
with $r=0.1$ and contribution from cosmic lensing. The red dashed
line (i.e. line 3) shows the square root of the average of the diagonal terms in the covariance matrix
$\left(\overline{\Delta{D}_{\ell}\Delta{D}_{\ell}}\right)^{1/2}$. The
blue solid (i.e. line 1) line shows the analytical result of $\langle D_{\ell}\rangle$, for a model with 
$r=0.1$ and contribution from cosmic lensing. The contributions two contributions
are shown separately with a magenta solid line (i.e. line 5) for the gravitational waves and a green solid line (i.e. line 6)
for the cosmic lensing. The blue dashed line (i.e. line 4) shows the analytical approximation (\ref{covariance-approximation}) for $\langle \Delta{D}_{\ell}\Delta{D}_{\ell}\rangle^{1/2}$.}
\label{fig9}
\end{figure}

In Fig.~\ref{fig9}, in order to verify (\ref{peudo-estimator}), we plot the left-hand side (solid blue line) of this equation for $C_{\ell}^{BB}$ a model with 
$r=0.1$ and contribution from cosmic lensing \cite{lensing}. For comparison, in this figure, we plot the average $\overline{D_{\ell}}$ over 1000 realizations for the same model (solid red line).
As expected the two lines are close to each other, being practically indistinguishable for multipoles $\ell\gtrsim20$. For comparison, in Fig.~\ref{fig9}, we also plot the individual 
contributions from gravitational waves (solid magenta line) and cosmic lensing (solid green line). 
Finally, in Fig.~\ref{fig9}, we plot the square root of the average over 1000 realizations of the diagonal 
terms in the covariance matrix $\left({\overline{\Delta{D}_{\ell}\Delta{D}_{\ell}}}\right)^{1/2}$ (dashed red line). 
In order to check the analytical approximation (\ref{covariance-approximation}), we also plot the diagonal term 
$\langle{{\Delta{D}_{\ell}\Delta{D}_{\ell}}}\rangle^{1/2}$ evaluated using the right side of expression (\ref{covariance-approximation}) (dashed blue line). As expected
the two curves for the covariance matrix practically coincide for large multipoles $\ell\gtrsim80$, which corresponds to the region of applicability 
of the approximation (\ref{covariance-approximation}).


\subsection{Unbiased estimators for $C_{\ell}^{BB}$\label{section4.2}}

Having constructed the pseudo estimator $D_{\ell}$, we are one step away from constructing an unbiased
estimator for the $B$-mode power spectrum $C_{\ell}^{BB}$. In this subsection, we shall discuss this construction.
Let us for the moment assume that the coupling matrix $M_{\ell \ell'}$ in (\ref{peudo-estimator}) is invertible. In this case, 
using relation (\ref{peudo-estimator}), one can immediately verify that the estimator defined as
\bea
D_{\ell}^{BB} = N_{\ell}^{-2} B_{\ell}^{-2} \sum_{\ell'} M^{-1}_{\ell \ell'}
{D}_{\ell'},
\ena
is an unbiased estimator of the power spectrum $C_{\ell}^{BB}$. In practice, this simple estimator
can be used in large sky surveys such as {Planck}, where $M_{\ell \ell'}$ is indeed invertible. 
However, in the case of small sky surveys, one cannot
construct this estimator since the matrix $M_{\ell \ell'}$ becomes singular. In this case it is possible to 
bin the pseudo estimator data into multipole bins, and construct an unbiased estimator for the binned
power spectrum. Following the analysis for temperature anisotropy \cite{hivon}, we build the so-called
full-sky CMB bandpowers ${\rm P}^{BB}_b$ as
\bea
{\rm P}^{BB}_b = \sum_{b'} M_{bb'}^{-1} \sum_{\ell} p_{b'\ell}
{D}_{\ell}, \label{bin-estimator}
\ena
where the subscript $b$ denotes the multipole bands. $p_{b\ell}$ is a
binning operator in $\ell$-space defined as
\bea
{p}_{b\ell}=\left\{
{\begin{array}{cc}
\frac{\ell(\ell+1)}{2\pi N^2_{\ell}(\ell_{\rm low}^{(b+1)}-\ell_{\rm low}^{(b)})},~&~{\rm if~\ell_{\rm low}^{(b)}\le\ell<\ell_{\rm low}^{(b+1)}
} \\ 
0~&~{\rm otherwise}
\end{array}
}
\right. 
\label{matrixP}.
\ena
The non-singular binned coupling matrix $M_{bb'}$ participating in (\ref{bin-estimator}) is
constructed from the coupling matrix $M_{\ell \ell'}$
\bea
M_{bb'} = \sum_{\ell} p_{b\ell} \sum_{\ell'} M_{\ell \ell'}
B_{\ell'}^2 q_{\ell'b'}.
\label{matrixM}
\ena
The function $B_{\ell'}^2$ takes into account the effects arising due to finite beam size and finite pixelization. In the above
expression, $q_{\ell b}$ is the reciprocal operator of
$p_{b\ell}$
\bea
 {q}_{\ell b}=\left\{
 {\begin{array}{cc}
 {\frac{2\pi N^2_{\ell}}{\ell(\ell+1)}},~&~{\rm if~\ell_{\rm low}^{(b)}\le\ell<\ell_{\rm low}^{(b+1)}
 } \\
 0~&~{\rm otherwise}
 \end{array}
 }
 \right. .
\ena
It is straightforward to verify that ${\rm P}^{BB}_b$ is an unbiased estimator
of the $B$-mode of polarization power spectrum $\ell(\ell+1)C_{\ell}^{BB}/2\pi$, i.e.
\bea
\langle{\rm P}^{BB}_b\rangle = \frac{\ell(\ell+1)}{2\pi}C_{\ell}^{BB}.
\nonumber
\ena
The covariance matrix of the bandpowers is related to the covariance matrix  
$\langle \Delta{D}_{\ell}\Delta{D}_{\ell'}\rangle$ in (\ref{covariancematrix}) by \cite{brown}
\bea
 \langle \Delta {\rm P}^{BB}_{b}\Delta{\rm
 P}^{BB}_{b'}\rangle = M_{bb_1}^{-1} p_{b_1\ell}
 \langle \Delta{D}_{\ell}\Delta{D}_{\ell'}\rangle
 (p_{b_2\ell'})^T (M_{b'b_2}^{-1})^T. 
\label{covarianceP}
\ena

In Fig.~\ref{fig10} we plot the value of the bandpower ${\rm P}^{BB}_b$ (red dots) averaged over 1000 realizations.
The realizations were generated for a model including contribution from gravitational waves with $r=0.1$ and cosmic lensing.
The multipole bins were chosen with $\Delta \ell=10$ for each bin. The analysis shows that (up to discrepancies that can 
be attributed to finite number of realizations) the average of the power spectrum estimators coincide with the theoretical (input) spectrum. 
The error bars $\left(\overline{\Delta {\rm P}^{BB}_{b}\Delta{\rm P}^{BB}_{b'}}\right)^{1/2}$ (red error bars) 
were calculated using (\ref{covarianceP}), with ensemble average replaced by an average over realizations. 
As can be expected, the error bars are large for the first three data points, due to the small sky coverage.
In addition to evaluating error bars, it was verified that the correlation between 
various multipole bins is QUIET weak (all of the correlation coefficients are smaller than $0.3$).
Note that, the correlation matrices and corresponding error bars calculated here 
do not include contribution from instrumental and astrophysical foreground noises. The left panel in Fig.~\ref{fig10} shows
the power spectrum estimation with both gravitational wave and cosmic lensing contributions included. The right panel shows the power spectrum 
estimation for the gravitational wave contribution alone. In the right panel, cosmic lensing serves as an effective noise for the detection of gravitational
waves (see next subsection for details). It can be seen that the gravitational wave
signal is larger than the corresponding error bars only in a range of 
multipoles $50\lesssim\ell\lesssim150$, peaked at $\ell\sim100$,
consistent with results in \cite{zz}. 

In order to quantify the detectability of gravitational wave signal, it is convenient to introduce
the total signal-to-noise ratio as follows
\bea
S/N=\sqrt{\sum_{bb'} \langle {\rm P}^{BB}_b({\rm gw})\rangle ({\rm Cov}^{-1})_{bb'}\langle{\rm P}^{BB}_{b'}({\rm gw})\rangle}, 
\label{snr}
\ena
where ${\rm Cov}_{bb'}\equiv\langle \Delta {\rm P}^{BB}_{b}\Delta{\rm P}^{BB}_{b'}\rangle$ is the covariance
matrix of the bandpower estimator (\ref{covarianceP}). For the example considered above with $r=0.1$, we find $S/N=8.26$.

It is important to emphasize that our
pseudo-$C_{\ell}$ estimator is QUIET different from the
pseudo-$C_{\ell}$ polarization estimators suggested in \cite{chon,challinor}, 
or an equivalent estimator suggested in \cite{quad-method}. 
In \cite{chon,challinor}, the unbiased estimators are constructed directly from the pseudo estimators
$\tilde{C}_{\ell}^{EE}$ and $\tilde{C}_{\ell}^{BB}$, both of which are a mixture of
electric and magnetic types of polarization. The resultant mixing increases the magnitude of the 
covariance matrix for the unbiased estimator, and becomes one of the main contaminations for the detection of
gravitational waves. In \cite{challinor}, the authors found that, for small sky surveys
covering one or two percent of the sky, the mixing contamination to the covariance matrix
of $B$-mode power spectrum estimator typically limits the tensor-to-scalar ratio that can be probed
to $r\gtrsim0.05$. On the other hand, the pseudo-$C_{\ell}$ method suggested in the present work explicitly
separates the electric and magnetic types of polarization up to very small numerical errors. For this reason, the
effects of mixing of electric and magnetic modes, which are completely removed (reduced to negligible levels)
in our case, do not put a limit on the ability to detect gravitational waves. This is the main advantage of our method,
and is the main motivation for this paper.

At the end of the subsection, we would like to point out that, if one proceeds to construct an unbiased estimator for
$C_{\ell}^{BB}$ using the $\mathcal{B}_{\rm rec}(\hat{\gamma})W(\hat{\gamma})$ field 
(instead of $\mathcal{B}_{\rm rec}(\hat{\gamma})$ field used above),
the resulting unbiased estimator will be equivalent to the ``pure B-mode" estimator defined in \cite{smith}. This has been discussed in Appendix \ref{appendixA3}. On the other hand,
if one constructs the unbiased estimator for $C_{\ell}^{BB}$ using $\tilde{\mathcal{B}}(\hat{\gamma})$ (instead of 
$\mathcal{B}_{\rm rec}(\hat{\gamma})$) adopting a top-hat window function (instead of $W(\hat{\gamma})$), one would
return to the $B$-mode estimator defined in \cite{kim}. As was pointed out in \cite{kim}, the resulting estimator 
suffers from large $EB$-mixing at the edge of the observed field.


\subsection{Information loss due to edge removal \label{section4.2a}}

As was emphasized in Sec. \ref{section3}, in practical calculations, the edge of the partial sky map has to be removed
in order to reduce numerical errors. The edge removal leads to the partial loss of information. In this subsection 
we study the impact of this information loss on the performance of $B$-mode of polarization power spectrum estimator
${\rm P}^{BB}_b$.

In order to study the performance of the estimator ${\rm P}^{BB}_b$ in an ideal case with no edge removal we perform the following
steps:
\begin{enumerate}
\item 
We generate 1000 full sky ($Q$, $U$) maps, for a cosmological model with $r=0.1$ and contribution from cosmic lensing.
For each of these maps we calculate the multipole coefficients $B_{\ell m}$ using (\ref{p-}), (\ref{multipoledecomposition}) and (\ref{eblm}).

\item
With the multipole coefficients $B_{\ell m}$ we construct the full sky map ${\mathcal{B}}(\hat{\gamma})$ using (\ref{mathcalelmblm2}).

\item
We construct the top-hat mask window function $w(\hat{\gamma})$ equal to unity for $\theta<\theta_0=20^o$ and zero otherwise.
We now construct the masked magnetic field ${\mathcal{B}}_{\rm rec}(\hat{\gamma})={\mathcal{B}}(\hat{\gamma})w(\hat{\gamma})$. The
masked field ${\mathcal{B}}_{\rm rec}(\hat{\gamma})$ constructed in this manner corresponds to a reconstructed magnetic field map
in an idealized case with no edge removal in the absence of numerical errors.

\item
Working with the field ${\mathcal{B}}_{\rm rec}(\hat{\gamma})$, following the steps outlined in Sec.~\ref{section4.1} and Sec.~\ref{section4.2},
we build the unbiased estimator ${\rm P}_{b}^{BB}$ and calculate its covariance matrix. The resulting estimator 
is equivalent to one that could be constructed in an ideal situation, without numerical errors, in which we could have worked without edge removal.
\end{enumerate}

\begin{figure}[t]
\centerline{\includegraphics[height=10cm,width=18cm]{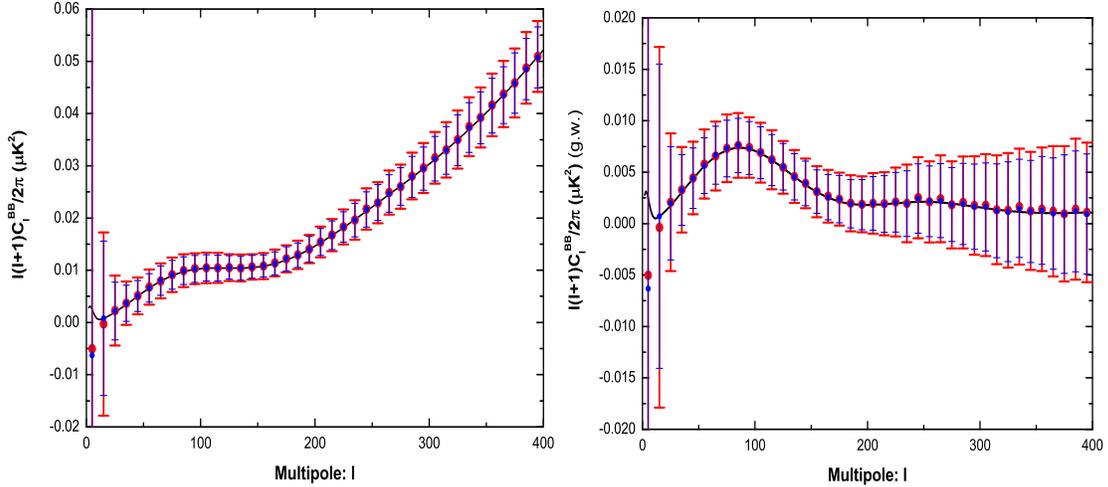}}
\caption{The averaged values (red larger dots) and error bars (red larger error bars) following from simulations
of the unbiased estimators of the power spectrum 
$\ell(\ell+1)C_{\ell}^{BB}({\rm total})/2\pi$ (left panel) 
and $\ell(\ell+1)C_{\ell}^{BB}({\rm gw})/2\pi$ (right panel). In both
panels, the black solid line denotes the theoretical values of the underlying power spectra. 
For comparison, in both panels, we plot the
averaged values (blue smaller dots) and simulated error bars (blue smaller error
bars) of the unbiased estimators for an ideal case without information loss
(see text for the details). In both the panels, we have considered a case
with no instrumental noise.} 
\label{fig10}
\end{figure}

\begin{figure}[t]
\centerline{\includegraphics[width=14cm]{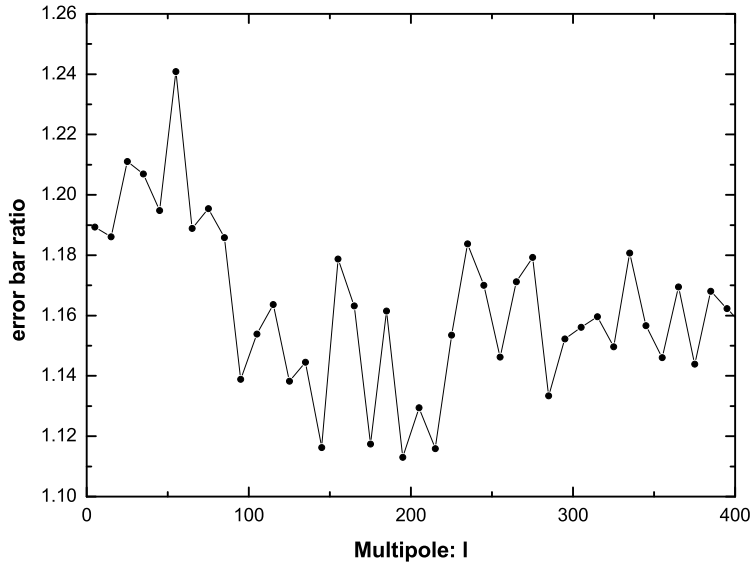}}
\caption{The ratio of practically achievable error bars (red error bars in
Fig. \ref{fig10}) to the corresponding error bars in an information lossless case (blue error bars in
Fig. \ref{fig10}), as a function of multipole $\ell$.}
\label{fig11}
\end{figure}

The resulting averaged value for the estimator (blue dots) and the corresponding error bars (blue error bars) are plotted in Fig.~\ref{fig10}.
Once again, we find that the average values of the estimators are practically coincident with the theoretical (input) values. 
In both the panels, the blue error bars are slightly smaller than the red ones for all multipole bins. The difference reflects the loss of
information due to edge removal. In Fig.~\ref{fig11}, we plot the ratio of the two error bars as a function of the multipole bin. This
ratio in almost everywhere less than $1.2$. The signal-to-noise ratio (\ref{snr}) calculated for the ideal is $S/N=9.59$, which is
less than $15\%$ higher than the practically relevant example considered in the previous subsection. The results of this section demonstrate
that the loss of information, gauged by the increase in the error bars of the spectral estimator, is sufficiently small $\lesssim15\%$.

\subsection{Power spectrum estimators in the presence of instrumental noise \label{section4.3}}

In the previous subsections we have considered a situation in which the magnetic type of polarization
was generated solely by gravitational waves and cosmic lensing. In realistic observations, in addition
to these two contributions, there are contaminating contributions from various other sources like
instrumental noise and astrophysical foregrounds. This said, it is reasonable to assume that, for an appropriate
choice of observed sky region, the astrophysical foregrounds are typically expected to be small
in comparison with instrumental noises \cite{bicep}. For this reason, we shall ignore the foreground contaminations, 
and restrict our analysis to the study of power spectrum estimators in the presence of only instrumental noises.

The pseudo estimator $D_{\ell}$ (\ref{pseudoDestimator}) has the following expectation in the presence of noise 
(compare with no noise case (\ref{peudo-estimator}))
\bea
\langle {D}_{\ell} \rangle = \sum_{\ell'}  M_{\ell \ell'}
N^2_{\ell'} B_{\ell'}^2 C_{\ell'}^{BB} + \langle{\mathcal{N}}_{\ell}^{BB}\rangle, \label{peudo-estimator-noise}
\ena
where ${\mathcal{N}}_{\ell}^{BB}$ is the pseudo estimator for the
full sky noise power spectrum $N_{\ell}^{BB}$. The expectation
value of this noise estimator is
\bea
 \langle{\mathcal{N}}_{\ell}^{BB}\rangle = \sum_{\ell'}  M_{\ell \ell'}
 N^2_{\ell'} N_{\ell'}^{BB} .  
\label{peudo-noise}
\ena
The presence of noise leads to a redefinition of the unbiased estimator ${\rm P}_{b}^{BB}$
\bea
{\rm P}^{BB}_b = \sum_{b'} M_{bb'}^{-1} \sum_{\ell} p_{b'\ell}
({D}_{\ell}-\langle{\mathcal{N}}_{\ell}^{BB}\rangle),
\label{bin-estimator-noise}
\ena
with matrices $p_{b'\ell}$ and $M_{bb'}$ given in (\ref{matrixP}) and (\ref{matrixM}).
The covariance matrix for this estimator has the form given by (\ref{covarianceP}), 
where $\langle\Delta D_{\ell}\Delta D_{\ell'}\rangle$ are calculated from the right side of (\ref{covariancematrix})
with $B_{\ell}^2C_{\ell}^{BB}$ terms replaced by $\left(B_{\ell}^2C_{\ell}^{BB}+N_{\ell}^{BB}\right)$.

The estimator ${\rm P}_{b}^{BB}$ defined in (\ref{bin-estimator-noise}) is an unbiased estimator
for the $B$-mode of polarization power spectrum $\ell(\ell+1)C_{\ell}^{BB}/2\pi$, where
$C_{\ell}^{BB}$ contains contribution from both gravitational waves (gw) and cosmic lensing (lens)
\bea
C_{\ell}^{BB} = C_{\ell}^{BB}({\rm gw})+C_{\ell}^{BB}({\rm lens}).
\nonumber
\ena
However, if we are primarily interested in detection of gravitational waves, we can treat the 
cosmic lensing contribution as effective noise, and define an unbiased estimator for the $B$-mode
power spectrum due to gravitational waves $\ell(\ell+1)C_{\ell}^{BB}({\rm gw})/2\pi$ as
\bea
{\rm P}^{BB}_b({\rm gw}) = \sum_{b'} M_{bb'}^{-1} \sum_{\ell} p_{b'\ell}
({D}_{\ell}-\langle{\tilde{\mathcal{N}}}_{\ell}^{BB}\rangle), 
\label{bin-estimator-noise-gw}
\ena
where the effective noise term $\langle\tilde{\mathcal{N}}_{\ell}^{BB}\rangle$ contains contribution from
instrumental noises and cosmic lensing
\bea
\langle{\tilde{\mathcal{N}}}_{\ell}^{BB}\rangle = \sum_{\ell'}  M_{\ell \ell'}
N^2_{\ell'} (B_{\ell'}^2 C_{\ell'}^{BB}({\rm len})+N_{\ell'}^{BB}) . 
\label{peudo-noise-gw}
\ena
The covariance matrix for this estimator is same as that calculated for estimator (\ref{bin-estimator-noise}).


\subsection{Expected performance of ground-based CMB experiments\label{section4.4}}

In this subsection we shall investigate the prospects of detecting the $B$-mode signature
from relic gravitational waves by two future ground based experiments, 
QUIET \cite{quiet} and POLARBEAR \cite{POLARBEAR}.

\begin{figure}[t]
\centerline{\includegraphics[height=12cm, width=18cm]{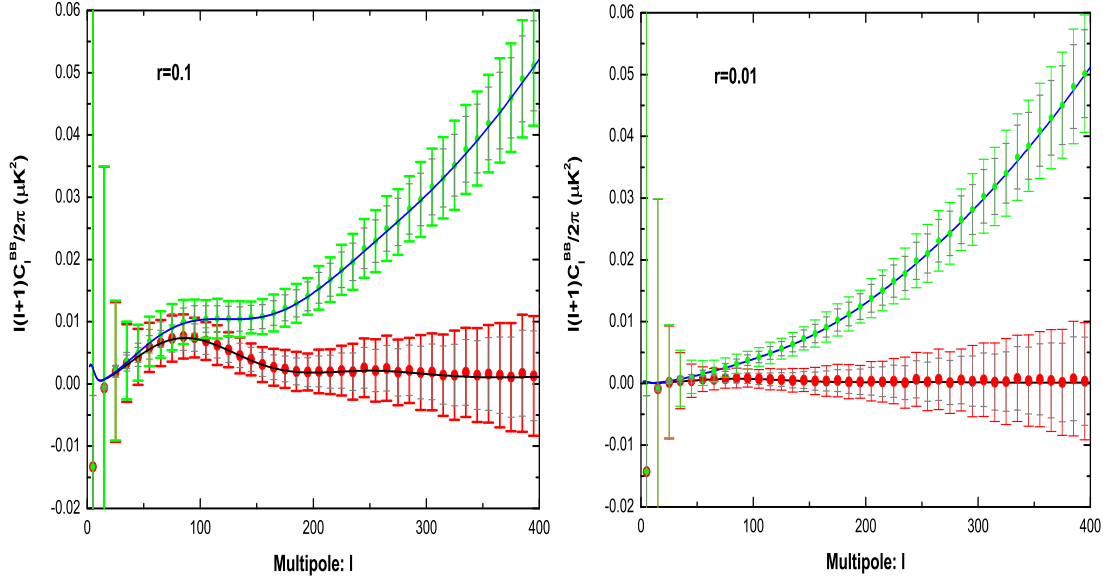}}
\caption{The averaged values and simulated error bars of the unbiased estimators for the power spectrum 
$\ell(\ell+1)C_{\ell}^{BB}/2\pi$ (green dots and green largers error bars) and 
$\ell(\ell+1)C_{\ell}^{BB}({\rm gw})/2\pi$ (red dots and red larger error bars). In both
panels, the solid lines denote the theoretical values for these power spectra. 
In this figure, we have considered the instrumental noise for QUIET experiment. 
The left panel shows the results for an input cosmological model with $r=0.1$, while the
right panel shows the results for an $r=0.01$ model. 
In both panels, the smaller error bars calculated using the analytical 
approximation (\ref{analytic-delta}) are plotted
in grey.}
\label{fig12}
\end{figure}

Let us firstly consider the QUIET experiment. We shall restrict our analysis to the 40GHz frequency channel. 
The FWHM for the Gaussian beam at this channel is $\theta_F=23'$,
and the expected instrumental noise is $N_{\ell}^{BB}=2.72\times10^{-7}\mu{\rm K}^2$ 
\cite{quiet} (see also \cite{zz}). We shall assume that
experiment will observe $f_{\rm sky}=3\%$ fraction of the sky, corresponding to $\theta_0=20^o$. following the steps
outlined in Sec.~\ref{section4.1} and Sec.~\ref{section4.2}, using the experimental characteristics for QUIET experiment,
we construct the unbiased estimators ${\rm P}^{BB}_b$ and ${\rm P}^{BB}_b(\rm{gw})$ and their covariance matrices for
1000 realizations with $r=0.1$ and $r=0.01$. The average values for the estimators and their corresponding error bars
are plotted in Fig.~\ref{fig12}, for $r=0.1$ (left panel) and $r=0.01$ (right panel). The error bars in this case are larger than
the error bars in Fig.~\ref{fig10} due to the inclusion of instrumental noises. The total signal-to-noise ratio in (\ref{snr})
is $S/N=7.05$ for $r=0.1$ model, and $S/N=1.25$ for the model with $r=0.01$. 

\begin{figure}[t]
\centerline{\includegraphics[width=14cm]{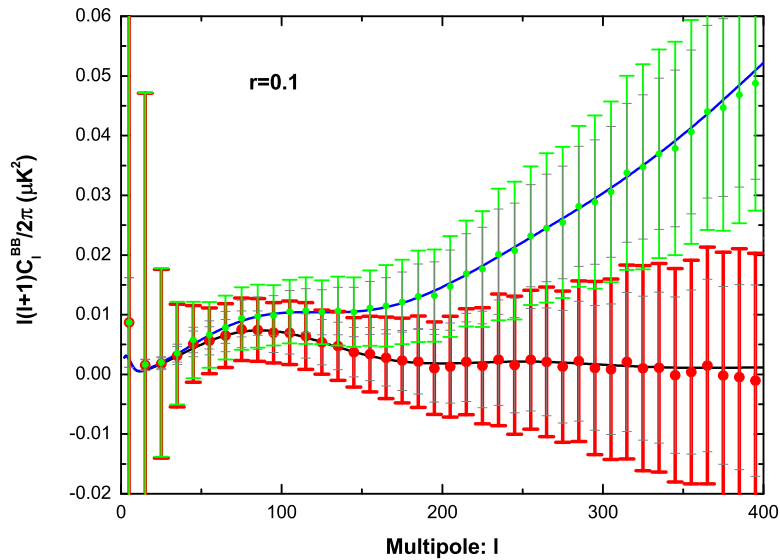}}
\caption{The results for POLARBEAR experiment. 
The averaged values and simulated error bars of the unbiased estimators for the power spectrum 
$\ell(\ell+1)C_{\ell}^{BB}/2\pi$ (green dots and green larger error bars) and 
$\ell(\ell+1)C_{\ell}^{BB}({\rm gw})/2\pi$ (red dots and red larger error bars), calculated for an input model with $r=0.1$.
The solid lines denote the theoretical values for these power spectra.
The smaller error bars calculated using the analytical 
approximation (\ref{analytic-delta}) are plotted in grey.} 
\label{fig13}
\end{figure}

We now turn to the POLARBEAR experiment. Once again, we restrict our study to the performance of the best frequency channel
at $150$GHz. The FWHM for the Gaussian beam is $\theta_F=4'$, and the expected instrumental noise is
$N_{\ell}^{BB}=4.22\times10^{-6}\mu$K$^2$ \cite{POLARBEAR}. As above, we assume $f_{\rm sky}=3\%$. In order to study the performance
of POLARBEAR, we simulate 1000 realizations of ($Q$, $U$) maps with $r=0.1$. Before proceeding to construct the power spectrum estimators,
one should notice that, in comparison with QUIET, the value of $\theta_F=4'$ for POLARBEAR is substantially smaller. Thus, in order to avoid leakage
from higher multipole electric type polarization, we firstly apply the `map smoothing' procedure outline in Appendix~\ref{appendixB}.
Following this, we construct the estimators ${\rm P}^{BB}_b$ and ${\rm P}^{BB}_b(\rm{gw})$ and their covariance matrices following the steps
explained in Sec.~\ref{section4.1} and Sec.~\ref{section4.2}. In Fig.~\ref{fig13} we plot the average values of the estimators and their error bars.
The error bars for POLARBEAR experiment are considerably larger than those in Fig.~\ref{fig12} (and Fig.~\ref{fig10}) due to larger instrumental noise
in comparison with QUIET. Finally, we calculate the signal-to-noise for the POLARBEAR experiment to be $S/N=4.31$ for
a model with $r=0.1$.

It is worth pointing out that, although in our estimations above we relied on the performances of a 
single best frequency channel for QUIET and POLARBEAR, these 
experiments will observe in several frequency channels. The combining of data from several frequency channels will
have an effect of reducing the total effective instrumental noise. In addition, these experiments could potentially observe larger portions of sky.
Both these points could potentially increase the detection ability of these experiments. On the other hand, one should remember that various
foregrounds \cite{foreground} and systematic errors \cite{systematics} would increase the effective noise, thereby reducing the detection ability.
One should remember these caveats, when looking at various signal-to-noise estimates, including the ones presented above.

At the end of this subsection we shall briefly discuss a widely used analytical approximation for signal-to-noise. In this approximation 
$S/N\propto\sqrt{f_{\rm sky}}$, where $f_{\rm sky}$ is the sky cut factor. This approximation follows from following considerations. In the case
of full sky coverage, one can construct an unbiased estimator $D_{\ell}^{XX}$ (where $X=T,E~{\rm or}~B$) for the various power spectra
$C_{\ell}^{XX}$ in a straightforward manner (see for example \cite{grishchuk,ttteee} for details). In this case the covariance matrix is
diagonal with 
\bea
\sqrt{\langle\Delta D_{\ell}^{XX}\Delta D_{\ell}^{XX}\rangle}=
\sqrt{\frac{2}{2\ell+1}}(C_{\ell}^{XX}+N_{\ell}^{XX}B_{\ell}^{-2}), 
\nonumber
\ena
with $\left(2\ell+1\right)$ in the denominator on the right side playing the role of number of degrees of freedom for a given multipole $\ell$.
The above expression was elegantly extrapolated for temperature anisotropy power spectrum estimator $D_{\ell}^{TT}$
to partial sky surveys in \cite{knox}. The author proposed to replace $\left(2\ell+1\right)$ with
the effective number of degrees of freedom $(2\ell+1)f_{\rm sky}$ in the above expression, to account for the loss of information that arrises
due to partial sky coverage. This simple consideration was extended to $B$-mode power spectrum estimator in \cite{jaffe}, and 
was further extended to account for multipole binning \cite{bowden}. These approximation lead to 
\bea
\langle\Delta {\rm P}_{b}^{BB}({\rm gw})\rangle = \sqrt{\frac{2}{(2\ell+1)\Delta\ell
f_{\rm sky}}} 
\left(\frac{\ell(\ell+1)}{2\pi}\right) 
\left(C_{\ell}^{BB}+N_{\ell}^{BB}B_{\ell}^{-2}\right), 
\label{analytic-delta}
\ena
with $\ell$ being the central multipole in each bin. In this approximation, the total signal-to-noise ratio for gravitational wave signal
in the $B$-mode of polarization takes the form
\bea
S/N= \sqrt{\sum_{b}\left(\frac{\langle{\rm P}_{b}^{BB}(\rm{gw})\rangle}{\langle\Delta {\rm P}_{b}^{BB}(\rm{gw})\rangle}\right)^2}.
\label{analytic-snr}
\ena
In order to guage the performance of this approximation, in Fig.~\ref{fig11} and Fig.~\ref{fig12},
we plot the error bars calculated using (\ref{analytic-delta}) (grey error bars). For this calculation we have set $f_{\rm sky}=0.024$
corresponding to an effective top-hat window function with $\theta_0=18^o$. One can see that, in both the figures, the analytical
approximation leads to smaller error bars than those obtained from numerical simulations. We use (\ref{analytic-snr}) to calculate
the analytical signal-to-noise ratio for the two considered experiments. The results for signal-to-noise ratio are summarized
in Table~\ref{table1}. It can be seen that the analytical approximation for signal-to-noise ratio (\ref{analytic-snr}) considerably overestimates
the detection ability, particularly for smaller values of actual $S/N$. Several works \cite{pixel,pixellikelihood} have pointed out that 
the analytical approximation (\ref{analytic-snr}) exaggerates the detection ability. However, these paper argued that the primary
reason for overestimation is due to the omission of possible contaminations from $EB$-mixing. However, our approach shows that the
analytical approximation (\ref{analytic-snr}) with an effective sky-cut factor also overvalues the $S/N$ in comparison with the case with no $EB$-mixing. One should therefore
use this approximation with caution \footnote{In \cite{hivon}, the authors found that formulae similar to in (\ref{analytic-delta}) and
(\ref{analytic-snr}) can over-valuate the detection of temperature anisotropies power spectrum. 
However, the authors argued that this overvaluation could be corrected by using a filter function when
building the unbiased estimators. We expect that a similar analysis can be applied in our method.}. At the same time,
it is very important to point out that this conclusion about overestimation is based on analysis of small sky coverage and the use of pseudo-$C_{\ell}$
estimators with the uniform weight function $\mathcal{W}(\hat{\gamma})$. In contrast, for large scale surveys \cite{hybrid} 
or the small scale surveys by using the pseudo-$C_{\ell}$ estimators with the optimal choice of the weight function $\mathcal{W}(\hat{\gamma})$ \cite{smith2}, the conclusion might change. 
Especially, for the large scale surveys and maximum likelihood estimators, the discussed analytical 
approximation may underestimate the true $S/N$, as was shown for temperature anisotropy in \cite{hybrid}.

\begin{table}
\caption{The total signal-to-noise $S/N$ for the gravitational waves signal in the $B$-mode of polarization
for the various cases considered in the text}
\begin{center}
\label{table1}
\begin{tabular}{|c|c|c|c|c|}
    \hline
     & ideal no noise case & QUIET noise& QUIET noise& POLARBEAR noise \\
     & $r=0.1$ & $r=0.1$ &$r=0.01$& $r=0.1$ \\
    \hline
    simulated $S/N$&8.26 &7.05 &1.25 &4.31 \\
   \hline
   analytical $S/N$&11.24 &10.76 &3.33 &7.15 \\
  \hline
\end{tabular}
\end{center}
\end{table}


\section{Conclusion \label{section5}}

In this paper we have proposed a new method to construct pure electric and magnetic type fields ${\mathcal{E}}(\hat{\gamma})$ and ${\mathcal{B}}(\hat{\gamma})$ from
polarization field given on an incomplete sky. Due to the differential definitions of these fields, we avoid the so-called $EB$-mixing problem. In practice when working with pixelized maps, residual leakages from various numerical errors require the removal of data from a narrow edge on the boundary of the observed sky. This leads to a minor loss of information in comparison with the idealized lossless case considered in Sec.~\ref{section4.2a}.

A major advantage of our approach is that the constructed fields ${\mathcal{E}}(\hat{\gamma})$ and ${\mathcal{B}}(\hat{\gamma})$ are scalar. For this reason, the various
techniques developed for the analysis of temperature anisotropy maps can be directly applied to these fields. As an important and motivating application, we discuss
the construction of an unbiased estimator for the $B$-mode power spectrum $C_{\ell}^{BB}$, using the pseudo-$C_{\ell}$ estimator approach. We find that our method
is computationally feasible even in the case of high resolution maps. In particular, it takes $2.5$ minutes on a laptop (2.4GHz processor and 2GB memory) to perform 
all of the calculations, including the the calculation of ${\mathcal{B}}(\hat{\gamma})$ in pixel space with $N_{\rm side}=512$, and the construction of unbiased estimators
for $C_{\ell}^{BB}$. 

With the help of the constructed unbiased estimator, we have investigated the ability to detect gravitational waves through the $B$-mode of polarization in CMB 
experiment covering $3\%$ of the sky. In the absence of instrumental noise, we find $S/N=8.26$ for a model with $r=0.1$. This value is $14\%$ smaller
than an idealized situation with no information loss. In the case of realistic experiments, the signal to noise reduces to $S/N=7.05$ for QUIET and $S/N=4.31$ for POLARBEAR.

In conclusion, we would like to point out that, a similar analysis can be applied to large sky surveys. In particular, for Planck satellite and the planned CMBPol experiment,
one can construct unbiased estimators for the polarization power spectra $C_{\ell}^{EE}$ and $C_{\ell}^{BB}$, by synthesizing the approach outlined in this paper together with
the hybrid estimator method suggested in \cite{hybrid}. We leave this task for future work.


~

\section*{Acknowledgements}
The authors appreciate help from E.~Hivon, L.~Cao and S.~Gupta in using the HEALPix package. The authors
thank L.~P.~Grishchuk for stimulating discussions. WZ is partially supported by Chinese NSF Grants No. 10703005, No.
10775119, and the Foundation for University Young Teaching Excellence of the Ministry of Education, Zhejiang Province.
In this paper, we have used the CAMB  package \cite{camb} and HEALPix package \cite{healpix}.


\appendix


\section{Numerical calculation of the correction term ${\rm ct}$ in pixel space \label{appendixA}}

In order to calculate the correction term ${\rm ct}$ in (\ref{Re-cts})
and (\ref{Im-cts}), one needs to be calculate the terms 
$(QW)_x$, $(QW)_y$, $(UW)_{x}$ and $(UW)_y$.
Below, we discuss the calculation of $(QW)_x$. 
The other terms are calculated in a similar manner. 

We expand the polarization fields $(Q+iU)W$ and $(Q-iU)W$ in terms of 
spin-weighted harmonics
\bea
 (Q(\hat{\gamma})\pm iU(\hat{\gamma}))W(\hat{\gamma})
 =\sum_{\ell m} \tilde{a}_{\pm2,\ell m}  ~_{\pm2}Y_{\ell
 m}(\hat{\gamma}).
 \label{Qw11}
\ena

It follows that
\bea
 QW(\hat{\gamma}) = 
 -\sum_{\ell m} \tilde{E}_{\ell m} X_{1,\ell m}(\hat{\gamma}) +
 i\tilde{B}_{\ell m} X_{2,\ell m}(\hat{\gamma}),
\ena
where 
\bea
\tilde{E}_{\ell m}\equiv-(\tilde{a}_{2,\ell m}+\tilde{a}_{-2,\ell m})/2,~~
\tilde{B}_{\ell m}\equiv-(\tilde{a}_{2,\ell m}-\tilde{a}_{-2,\ell
m})/2i ,
\nonumber \\
X_{1,\ell m}=(~_2Y_{\ell m}+ ~_{-2}Y_{\ell m})/2,~~X_{2,\ell m}=(~_2Y_{\ell m}- ~_{-2}Y_{\ell m})/2.
\nonumber
\ena
The quantity $(QW)_x$ can be numerically calculated using
\bea
 (QW)_x\equiv {\partial (QW)}/{\partial\theta} = -\sum_{\ell m} \tilde{E}_{\ell m} (\partial X_{1,\ell m}/\partial \theta) +
 i\tilde{B}_{\ell m} (\partial X_{2,\ell m}/\partial \theta).
 \label{Qw22}
\ena
Thus, using the expansion coefficients in (\ref{Qw11}) and expression (\ref{Qw22})
one can calculate the quantity $(QW)_x$ in terms of quantities $QW$, $UW$ and
functions  $(\partial X_{n,\ell m}/\partial \theta)$. We would like mention here that
in the HEALPix version 1.23, the subroutine \emph{alm\_map\_template.f90} had a bug,
that led to erroneous results for $(QW)_x$ and $(UW)_x$ \cite{healpix1.23}. This
problem has been fixed in the latest HEALPix version 1.24.

In the present paper we use a simple analytical form (\ref{window}) for 
the mask window function $W(\hat{\gamma})$. For this window function, the various derivatives, 
such as $W_x(\hat{\gamma})$ and $W_{xx}(\hat{\gamma})$, can be calculated analytically.
However, in practical situations, the window function does not have such a simple form
(see for instant \cite{wmapmask}). For this reason, one would need to calculate the various
derivative terms, $W_x$, $W_y$, $W_{xx}$, $W_{yy}$ and $W_{xy}$, numerically. This
can be done in the following way. One firstly defines the multiple expansion coefficients
$W_{\ell m}$ in the standard way
\bea 
W_{\ell m}\equiv \int W(\hat{\gamma}) Y^*_{\ell m}
(\hat{\gamma}) d\hat{\gamma}. 
\nonumber
\ena 
Following this, one calculates
\bea
W_x(\hat{\gamma})\equiv \frac{\partial W}{\partial
\theta}&=&\sum_{\ell m}
W_{\ell m} \left(\frac{\partial }{\partial \theta}Y_{\ell m}(\hat{\gamma})\right) \nonumber \\
&=&\sum_{\ell m} W_{\ell m} \left(\frac{\ell}{\tan\theta}Y_{\ell
m}(\hat{\gamma})-\frac{1}{\sin\theta}\sqrt{\frac{2\ell+1}{2\ell-1}(\ell^2-m^2)}Y_{\ell-1
m}(\hat{\gamma})\right). 
\nonumber
\ena 
The other quantities can be calculated in an anologous manner. It is important to point out that
the steps mentioned above can be realized in a straightforward manner using the 
\emph{anafast} and \emph{synfast} routines in the HEALPix package.


\section{Construction of magnetic map $\mathcal{B}_{\rm rec}(\hat{\gamma})$ from simulated polarization maps\label{appendixA2}}

In this appendix, we outline the steps which were used to simulate the polarization maps and construct the pure magnetic map $\mathcal{B}_{\rm rec}(\hat{\gamma})$.

\begin{enumerate}

\item 
We generate the mask window function $W(\hat{\gamma})$ using (\ref{window}) in pixel space using the standard pixelization scheme used in HEALPix with 
$N_{\rm side}=512$ (or $N_{\rm side}=1024$ in the example in Sec.~\ref{section3.2}).

\item
Using {\it synfast} HEALPix routine, we generate full sky ($Q(\hat{\gamma})$, $U(\hat{\gamma})$) maps with $N_{\rm side}=512~{\rm or}~1024$, using
cosmological parameters (\ref{background}) and appropriate value of tensor-to-scalar ratio $r$ as input. Using the window function
$W(\hat{\gamma})$, we build the masked ($\tilde{Q}(\hat{\gamma})$, $\tilde{U}(\hat{\gamma})$) maps 
(where $\tilde{Q}=QW$ and $\tilde{U}=UW$).

\item
Using  {\it anafast} HEALPix routine, we calculate the coefficients ($\tilde{{E}}_{\ell m}$, $\tilde{{B}}_{\ell m}$). The field $\tilde{\mathcal{B}}(\hat{\gamma})$ is
calculated from $\tilde{{B}}_{\ell m}$ according to (\ref{pseudo-epsilonbeta}) using {\it synfast} routine.

\item
With coefficients ($\tilde{{E}}_{\ell m}$, $\tilde{{B}}_{\ell m}$) we construct the fields 
$QW(\hat{\gamma})$, $UW(\hat{\gamma})$, $(UW)_x(\hat{\gamma})$ and $(QW)_y(\hat{\gamma})$, using the $5^{\rm th}$ option in {\it synfast} routine.

\item
Using the fields $UW(\hat{\gamma})$, $(UW)_x(\hat{\gamma})$ and $(QW)_y(\hat{\gamma})$ constructed in the previous step and analytical expressions 
for $W$, $W_x$ and $W_{xx}$ (derived by differentiating (\ref{window})), we calculate ${\rm Im}({\rm ct})$ in (\ref{imagine-cts}).

\item
The pure magnetic field $\mathcal{B}_{\rm rec}(\hat{\gamma})$ is now constructed from $\tilde{\mathcal{B}}$, $W$ and ${\rm Im}({\rm ct})$ using (\ref{main3}).
The pure magnetic field $\mathcal{B}_{\rm rec}(\hat{\gamma})$ is truncated at the edges in order to remove residual
leakages associated with numerical errors.

\end{enumerate}


\section{Pseudo Estimators for a special choice of weight function ${\mathcal{W}}(\hat{\gamma})=W(\hat{\gamma})$\label{appendixA3}}

In Sec. \ref{section4.1} it was pointed out that, in principle one can construct pseudo estimators of the power spectrum by adopting an arbitrary weight function $\mathcal{W}(\hat{\gamma})$ in (\ref{almrec-weighted}). Above, in the main text, we have focused on a specific case corresponding to a uniform weight function $\mathcal{W}(\hat{\gamma})=1$ . This choice is nearly optimal for high multipoles. However, this choice becomes sub-optimal at lower multipoles \cite{smith2}. In this appendix we study another possible choice for the weight function, namely  $\mathcal{W}(\hat{\gamma})=W(\hat{\gamma})$, where $W(\hat{\gamma})$ is the mask window function in (\ref{windowfunction}) with $\theta_0=20^o$ and $\theta_1=10^o$.  Note that the function $W(\hat{\gamma})$ is the same window function that was used for constructing $\mathcal{B}_{\rm rec} (\hat{\gamma})$. With this choice, the resulting pseudo estimator is equivalent to the pure $B$-mode estimator studied in \cite{smith}.

The construction of the pseudo estimators and the corresponding unbiased estimators follows closely the discussion in Sec. \ref{section4}. The only difference is that the definition of coefficients $a_{\ell m}$ in (\ref{almrec}) are modified to
\bea
a_{\ell m} = \int d\hat{\gamma}~ {\mathcal{B}}_{\rm rec}(\hat{\gamma}) W(\hat{\gamma}) Y^*_{\ell m}(\hat{\gamma}),
\label{almrec-smith}
\ena
and the quantities ${w'}(\hat{\gamma})$ and ${w}'_{\ell}$ in (\ref{klmlm}) and (\ref{mll}) would now be replaced by $W$ and its power spectrum, respectively.

\begin{figure}[t]
\centerline{\includegraphics[height=10cm,width=18cm]{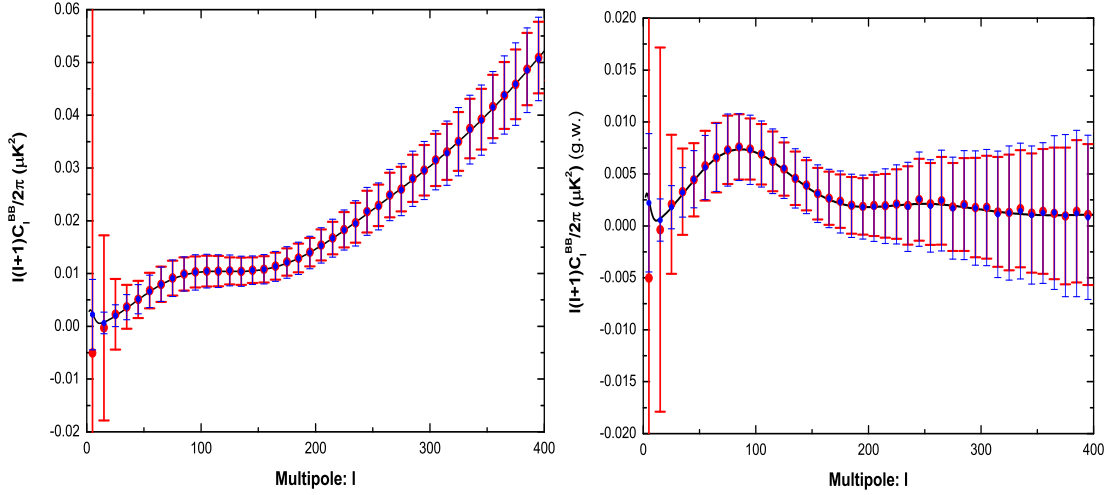}}
\caption{The averaged values (blue smaller and red larger dots) and error bars (blue smaller and red larger error bars) following from simulations of the unbiased estimators of the power spectrum $\ell(\ell+1)C_{\ell}^{BB}({\rm total})/2\pi$ (left panel) 
and $\ell(\ell+1)C_{\ell}^{BB}({\rm gw})/2\pi$ (right panel). In both
panels, the black solid line denotes the theoretical values of the underlying power spectra. 
The blue smaller dots and error bars denote the result by adopting the weight function $\mathcal{W}(\hat{\gamma})=W(\hat{\gamma})$, and
the red larger dots and error bars denote the result by adopting a uniform weight function $\mathcal{W}(\hat{\gamma})$. Note that, in both panels, we have considered a case
with no instrumental noise. The red larger dots and error bars are identical to those in Fig. \ref{fig10}.} 
\label{figsmith}
\end{figure}

In Fig. \ref{figsmith}, we plot the unbiased estimators for the power spectra $\ell(\ell+1)C_{\ell}^{BB}({\rm total})/2\pi$ (left panel) and $\ell(\ell+1)C_{\ell}^{BB}({\rm gw})/2\pi$ (right panel) together with the corresponding error bars (thin blue error bars). It can be seen that, in comparison with the estimators in the case of a uniform weight function, the error bars of the new estimators are larger at high multipoles, but are smaller at low multipoles. This result is consistent with findings in \cite{smith2} that the optimal weight functions for high multipoles tend to the top-hat function. On the other hand, for low multipoles, the optimal weight function tend to smooth out (see the right panel of Fig.2 in \cite{smith2} for a concrete example).

\section{Smoothing the polarization maps \label{appendixB}}

The high value of power spectrum $C_{\ell}^{{\mathcal{E}}{\mathcal{E}}}$ of
the electric component at large values of multipoles (due to the presence of $N^2_{\ell}$ factor)
leads to substantial leakage of power into the reconstructed pure magnetic field 
$\mathcal{B}_{\rm rec}$. This leakage seeps through to low multipoles playing a role of
residual effective noise. In order to reduce this contamination, below we introduce 
a map smoothing procedure for polarization maps. The idea behind this method is 
similar to the `prewhitening' method suggested in \cite{prewhiten}.

The smoothing procedure is simple and straightforward in the case of full sky coverage.
Given the ($Q$, $U$) polarization maps on a full sky, one can calculate the multipole coefficients
$E_{\ell m}$ and $B_{\ell m}$ using (\ref{p+-1})-(\ref{eblm}). In order to smooth the polarization
maps we firstly use a damping function to modify the multipole coefficients
\bea
 E'_{\ell m}\equiv E_{\ell m}e^{-\frac{1}{2}\frac{\ell^{2}\theta_F^2}{8\ln
 2}},~~
  B'_{\ell m}\equiv B_{\ell m}e^{-\frac{1}{2}\frac{\ell^{2}\theta_F^2}{8\ln
 2}}.
\ena
Following this, we reconstruct the smoothed polarization fields $Q'$ and $U'$
using the modified coefficients $E'_{\ell m}$ and $B'_{\ell m}$ in the standard way.
The reconstructed maps can be thought of as the the result of observing the original
($Q$, $U$) polarization field in an experiment with FWHM of the Gaussian beam equal
to $\theta_F$. Overall, the smoothing has effect of exponentially damping the power
in high multipoles.

We can extend the smoothing procedure to the case of partial sky coverage. Given
the ($Q$, $U$) polarization maps on a partial sky, we calculate the coefficients
$\tilde{E}_{\ell m}$ and $\tilde{B}_{\ell m}$. These coefficients are smoothed in analogy
with full sky case
\bea
 \tilde{E'}_{\ell m}\equiv \tilde{E}_{\ell m}e^{-\frac{1}{2}\frac{\ell^{2}\theta_F^2}{8\ln
 2}},~~
  \tilde{B'}_{\ell m}\equiv \tilde{B}_{\ell m}e^{-\frac{1}{2}\frac{\ell^{2}\theta_F^2}{8\ln
 2}}.\label{smooth}
\ena
The smoothed polarization maps ($Q'$, $U'$) are reconstructed from the modified
multipole coefficients $\tilde{E'}_{\ell m}$ and $\tilde{B'}_{\ell m}$.

It is important to point out that, in the case of partial sky coverage, the smoothing
procedure outlined above introduces mixture of electric and magnetic polarizations.
In particular, even if the original ($Q$, $U$) did not contain magnetic type of polarization,
the smoothed map ($Q'$, $U'$) would contain it. We have verified numerically that in practically
interesting cases the resulting mixture is very small, and would not significantly affect the ability to 
detect gravitational waves.

In order to verify the small of the resulting mixing, we have performed the following
calculation. Using an input model with no $B$-mode of polarization (i.e.~$C_{\ell}^{BB}=0$) 
and $\theta_F=30'$ we generated a full sky ($Q$, $U$) map. Following this, we truncate the map
to keep the data from only the northern hemisphere. Using the procedure outlined in 
Sec.~\ref{section2} and Sec.~\ref{section3} we construct the field $\mathcal{B}_{\rm rec}$,
which is expected to be equal to zero except for the residual leakage. The psedo power spectrum $D_{\ell}$
for this field (red line) is plotted in Fig.~\ref{fig14}. Following this, for the same input model, we generate the ($Q$, $U$) map
with $\theta_F=10'$, once again restricting the data to just the northern hemisphere. We now smooth this map with $\theta_F=30'$
using the {\it anafast}, {\it alteralm} and {\it synfast} HEALPix routines. We construct $\mathcal{B}_{\rm rec}'$ from the smoothed ($Q'$, $U'$) map
and plot the corresponding pseudo power spectrum $D_{\ell}'$ (green line) in Fig.~\ref{fig14}.
The difference between the two spectra $D_{\ell}$ and $D_{\ell}'$ can be interpreted as the result of mixing introduced
by map smoothing (blue line in Fig.~\ref{fig14}). It can be seen that, the mixing due to smoothing is QUIET small, comparable
to leakage due to numerical errors, at all the relevant multipoles. The power spectrum of the leakage due to smoothing
peaks at $\ell\sim400$. It can be completely neglected at low multipoles ($\ell\lesssim50$).

Below we give a heuristic argument to understand these results. The two sets of multipole coefficients 
($\tilde{E}_{\ell m}$, $\tilde{B}_{\ell m}$) and (${E}_{\ell m}$, ${B}_{\ell m}$) are related by the following
expression (see for instant \cite{challinor}) 
\bea 
\tilde{E}_{\ell m} + i \tilde{B}_{\ell m} = 
\sum_{\ell' m'} ~_{2}I_{(\ell m)(\ell' m')}
[E_{\ell' m'}+i B_{\ell', m'}], 
\ena 
where $~_{2}I_{(\ell m)(\ell' m')}$ is the coupling matrix, which depends only on the mask window
function. From this relation, it formally follows that
\bea
{E_{\ell m}} + i {B_{\ell m}} = \sum_{\ell' m'} ({~_{2}I}^{-1})_{(\ell m)(\ell'
m')} [\tilde{E}_{\ell' m'}+i \tilde{B}_{\ell', m'}]. 
\label{b4}
\ena
For the smoothed multipole coefficients one has
\bea 
{E'_{\ell m}} + i{B'_{\ell m}}&\equiv& 
\sum_{\ell' m'} ({~_{2}I}^{-1})_{(\ell m)(\ell' m')} [\tilde{E'}_{\ell' m'}+i \tilde{B'}_{\ell', m'}] \nonumber\\
&=& \sum_{\ell' m'} ({~_{2}I}^{-1})_{(\ell m)(\ell' m')}
[\tilde{E}_{\ell' m'}+i \tilde{B}_{\ell', m'}]
e^{-\frac{1}{2}\frac{\ell'^{2}\theta_F^2}{8\ln 2}}. 
\nonumber
\ena
Since the coupling matrix $ {~_{2}I}_{(\ell m)(\ell' m')}$ is sharply peaked at $\ell=\ell'$ \cite{challinor},
the above expression can be approximated by
\bea 
{E'_{\ell m}} + i {B'_{\ell m}} &\approx& 
\sum_{\ell' m'} ({~_{2}I}^{-1})_{(\ell m)(\ell' m')} [\tilde{E}_{\ell' m'}+i
\tilde{B}_{\ell', m'}]
e^{-\frac{1}{2}\frac{\ell^{2}\theta_F^2}{8\ln 2}}
\nonumber\\&=&(E_{\ell m}} + i {B_{\ell
m})e^{-\frac{1}{2}\frac{\ell^{2}\theta_F^2}{8\ln 2}}.
\nonumber
\ena
It therefore follows that
\bea 
{E'_{\ell m}} \approx E_{\ell m}
e^{-\frac{1}{2}\frac{\ell^{2}\theta_F^2}{8\ln 2}},~~{B'_{\ell m}}
\approx B_{\ell m} e^{-\frac{1}{2}\frac{\ell^{2}\theta_F^2}{8\ln
2}}. 
\ena
Thus, in this approximation, there is no mixing introduced by smoothing. Moreover,
$e^{-\frac{1}{2}\frac{\ell'^{2}\theta_F^2}{8\ln 2}}\approx 1$ for $\ell'\ll1/\theta_F$, 
leading to ${E'_{\ell m}} \approx E_{\ell m}$, ${B'_{\ell m}} \approx B_{\ell m}$ for low multipoles. 
In reality the coupling matrix $ {~_{2}I}_{(\ell m)(\ell' m')}$
is often non-invertible, and therefore (\ref{b4}) might not be valid. However, the resulting leakage 
is still small, and completely negligible for $\ell\ll1/\theta_F$.





\begin{figure}[t]
\centerline{\includegraphics[width=14cm]{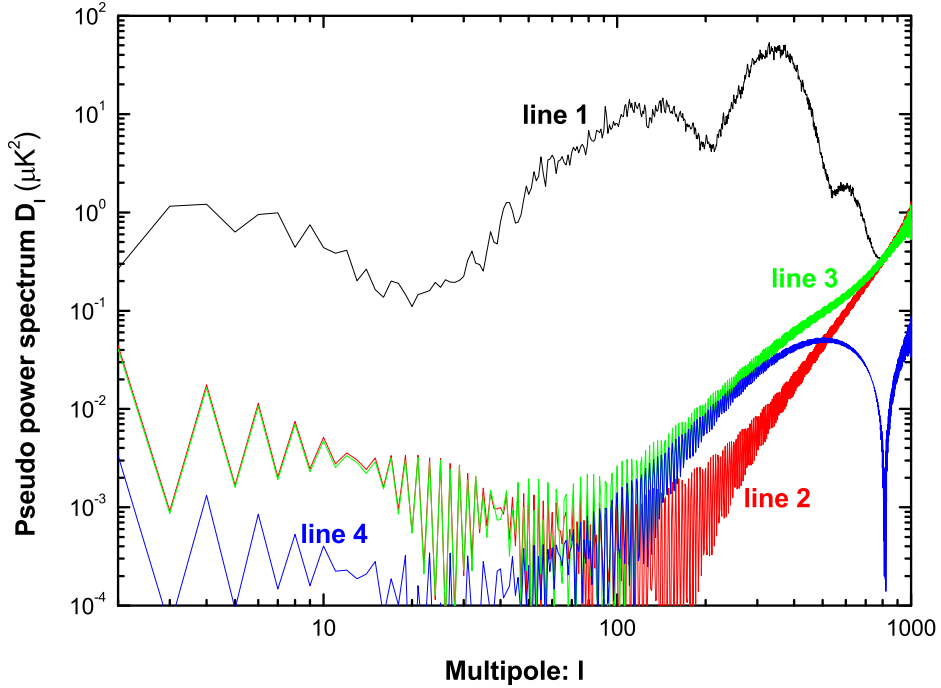}}
\caption{
The red line (i.e. line 2) shows the pseudo power spectrum $D_{\ell}$ calculated for a map with $\theta_F=30'$.
The green line (i.e. line 3) shows $D_{\ell}$ constructed for a map which was, smoothed from $\theta_F=10'$ to $\theta_F=30'$. 
The blue line (i.e. line 4) shows the difference between the green and the red lines, 
and corresponds to the leakage from ${\mathcal{E}}$ polarization to ${\mathcal{B}}$ polarization due to the map smoothing. 
For comparison, the black line (i.e. line 1) shows the pseudo power spectrum
$D_{\ell}^{\tilde{{\mathcal{B}}}\tilde{{\mathcal{B}}}}$.
Note that, the black line (i.e. line 1) and red line (i.e. line 2) are identical to those in Fig.~\ref{fig3}.} 
\label{fig14}
\end{figure}


\end{document}